\documentclass[aps,prl,twocolumn,showpacs,notitlepage,floatfix,longbibliography,superscriptaddress,10pt]{revtex4-1}

\usepackage{amssymb}
\usepackage{amsmath}
\usepackage{graphicx}
\usepackage{braket}
\usepackage[colorlinks=true,linkcolor=blue,citecolor=blue,filecolor=blue,urlcolor=blue,pdfstartview=FitH,breaklinks=true]{hyperref}
\usepackage{xcolor}
\usepackage{ulem}
\usepackage{diagbox} 

\newcommand{\pcsadd}{Center for Theoretical Physics of Complex Systems, Institute for Basic Science (IBS), Daejeon, Korea, 34126}
\newcommand{\ustadd}{Basic Science Program, Korea University of Science and Technology (UST), Daejeon 34113, Republic of Korea}
\newcommand{\nziasadd}{Centre for Theoretical Chemistry and Physics, The New Zealand Institute for Advanced Study, Massey University Auckland, Auckland, New Zealand}

\newcommand{\iscadd}{Institute for Complex Systems, National Research Council (ISC-CNR), Via dei Taurini 19, 00185 Rome, Italy}
\newcommand{\Qstaradd}{Centre for Quantum Technologies, National University of Singapore, 3 Science Drive 2, Singapore 117543}

\newcommand{\mh}{\mathcal{H}}
\newcommand{\mhsp}{\ensuremath{\hat{\mh}_\text{sp}}}

\begin{document}

\title{Flat band fine-tuning and its photonic applications}
%Designing many-body localized quantum lattices\\
%\underline{Buzz-words}: interaction, caging, many-body dynamics, dispersionless network

\author{Carlo Danieli}
\affiliation{\iscadd}

\author{Alexei Andreanov}
\affiliation{\pcsadd}
\affiliation{\ustadd}

\author{Daniel Leykam}
\affiliation{\Qstaradd}

\author{Sergej Flach}
\affiliation{\pcsadd}
\affiliation{\ustadd}
\affiliation{\nziasadd}

\date{\today}

\begin{abstract}
    Flat bands -- single-particle energy bands -- in tight-binding networks have attracted attention due to the presence of macroscopic degeneracies and their extreme sensitivity to perturbations.
    This makes them natural candidates for emerging exotic phases and unconventional orders.
    The challenging part however is to construct flat band networks, whose existence relies on symmetries and fine-tuning.
    In this review we consider the recently proposed systematic ways to construct flat band networks based on symmetries or fine-tuning.
    We then discuss how the fine-tuning constructions can be further extended, adapted or exploited in presence of perturbations, both single-particle and many-body.
    This strategy has lead to the discovery of non-perturbative metal-insulator transitions, fractal phases, nonlinear and quantum caging and many-body nonergodic quantum models.
    We discuss what implications these results may have for the design of fine-tuned nanophotonic systems including photonic crystals, nanocavities, and metasurfaces.
\end{abstract}

\maketitle

\section{Introduction}
\label{sec:I}

\begin{figure}[h!]
    \centering \includegraphics[width=0.925\columnwidth]{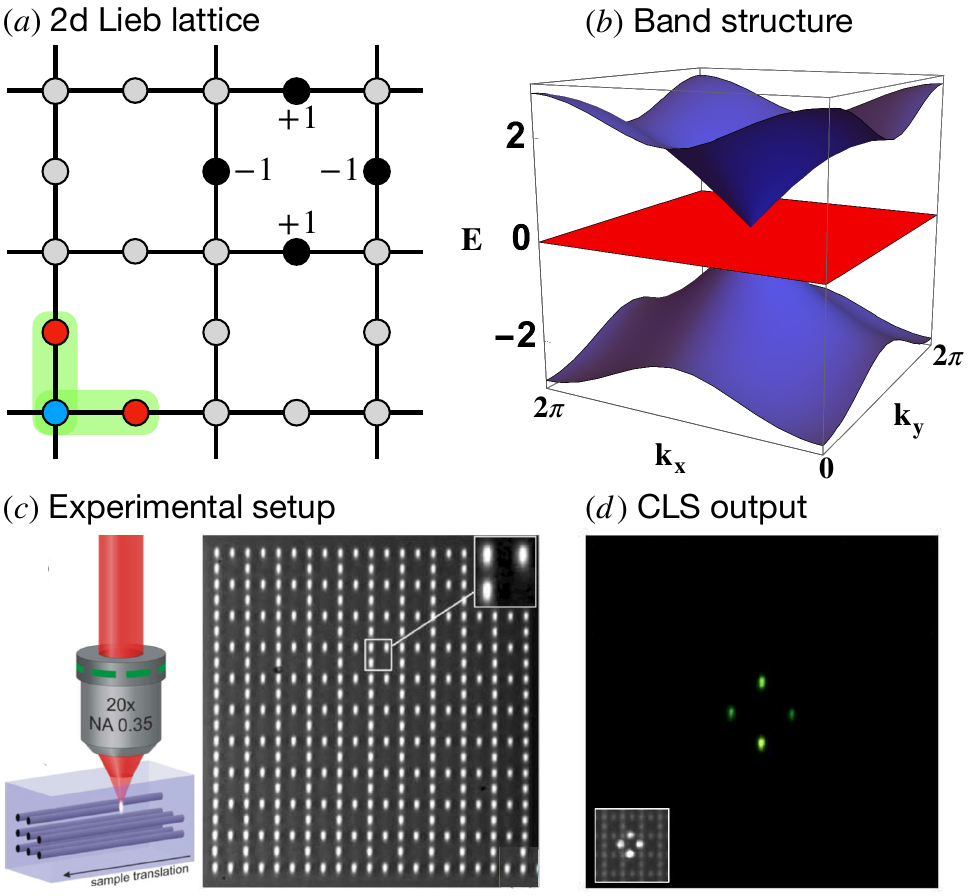}
    \caption{
        (a) 2d Lieb lattice and the location of the CLS. 
        Solid circles show the CLS profile with given amplitudes and alternating phases (indicated by the signs).
        The green shaded area indicates the unit cell, with the majority sublattice (red) and the minority sublattice (blue).
        (b) Band structure with the clearly visible flat band (red) between two dispersive bands (blue). 
        (c) Femtosecond-laser writing technique on a silica glass wafer and microscope image at the output facet of a Lieb lattice for white light propagation, adapted from Ref.~\cite{vicencio2015observation}
        (d) Output experimental pattern for an initial CLS excitation. The inset shows the input intensity profile~\cite{vicencio2015observation}.
    }
    \label{fig:lieb}
\end{figure}

Flat bands (FB) are strictly dispersionless bands embedded in a band structure of waves which in general can propagate through a spatially periodic medium~\cite{derzhko2015strongly,leykam2018artificial,leykam2018perspective}.
For finite range hopping they are usually accompanied with the existence of compact localized eigenstates (CLS). 
In Fig.~\ref{fig:lieb} (a) the two-dimensional Lieb lattice is shown with its CLS occupying four sites, as well as the corresponding band structure with two dispersive and flat bands in Fig.~\ref{fig:lieb}(b).
The CLS is an exact eigenstate of the system.
Despite existing network connections to unoccupied sites the CLS resists any expansion due to destructive interference.
Photonic realizations of this CLS and flat band example were reported by Vicencio et al~\cite{vicencio2015observation} and Mukherjee et al~\cite{mukherjee2015observation}.
The authors used a femtosecond-laser writing on silica glass wavers to generate the Lieb lattice (Fig.~\ref{fig:lieb}(c), and successfully propagated light through the resulting photonic waveguide network in a compact localized state as shown in Fig.~\ref{fig:lieb}(d).

%   Figure 1

The fascination in flat bands derives from the fact that a flat band results in macroscopic degeneracy and absence of transport, which - once broken, even slightly by tiny perturbations - result in novel transporting phases whose properties depend on the type of perturbations added. 
The perturbations hence become the game maker, and that is what turns flat bands into interesting starting points for switching between different phases of matter through tiny changes of tiny perturbations. 
Consequently, perturbation terms have to be controlled through meticulous fine-tuning. 
The same also applies to flat band lattices themselves, as they are part of manifolds and different flat band lattices respond differently to the same perturbation. 
Fine-tuning within these manifolds is therefore a desirable scheme which will allow for the above-mentioned switching.

The still-young field of flat band physics went through a few evolution cycles already. 
It probably first manifested through an observation of Bill Sutherland that compact localized eigenstates (CLS) are not a prerogative of tight-binding models of quasicrystals~\cite{kohmoto1986electronic1}, but persist also in systems with strict discrete translational invariance~\cite{sutherland1986localization}.
Then Elliott Lieb used the Lieb lattice and its flat band for exciting high \(T_c\) superconductivity physics of charge transport in Co-O\(_2\) planes on one side~\cite{lieb1989two}, 
but on the other side firmly introduced the concept of chiral symmetry protected FBs, see Fig.~\ref{fig:lieb}.
The first cycle was completed by Mielke~\cite{mielke1991ferromagnetism} and Tasaki~\cite{tasaki1992ferromagnetism}, with Mielke offering a flat band generator using line graphs, 
and Tasaki showing that the Lieb lattice FB can be preserved under specific well chosen fine-tuned perturbations of the Lieb lattice. 
While both aimed at and described novel features of ground and excited states of an electronic many-body interacting system, we stress here their contribution to the first evolution cycle of the single particle FB physics field.

\begin{figure}[h!]
    \centering
    \includegraphics[width=0.925\columnwidth]{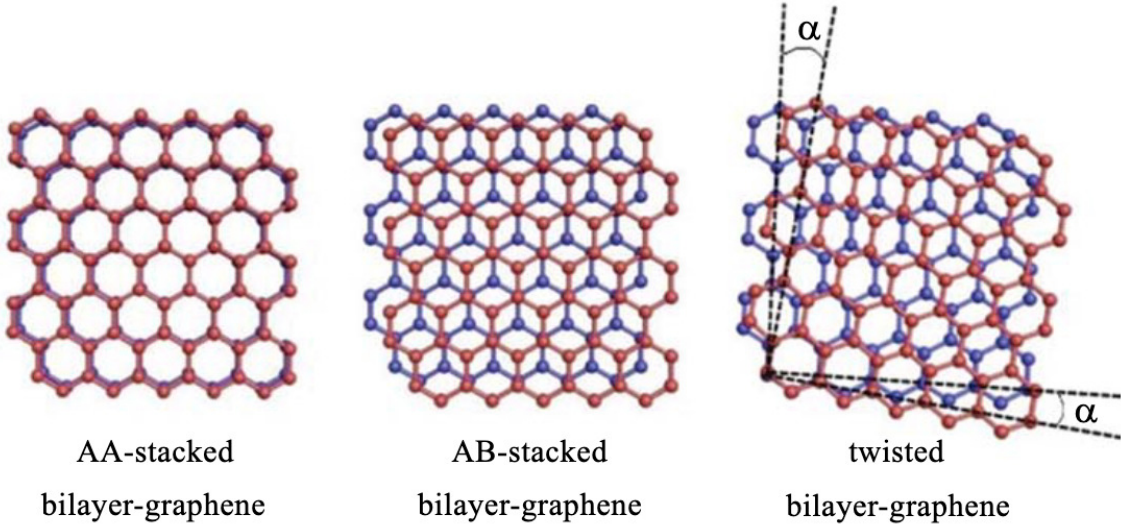}
    \caption{
        The structure diagram of stacked and twisted bilayer graphene. 
        Adapted from Ref.~\cite{huang2019bi}.
    }
    \label{fig:bilayer}
\end{figure}

During the second cycle of about 20 years an impressive amount of publications - mainly theoretical - was added targeting condensed matter-related topics and questions~\cite{derzhko2015strongly}.
Many aspects and details of FB properties were reported, and a number of qualitatively different properties were identified during this second cycle hinting at the existence of different classes of FBs.
However, it was far from being clear how to systematically search for new flat band models even at the level of noninteracting single particle physics.
This is mainly due to the condensed matter material science platform, which does not allow to easily prepare materials with fine-tuned electronic properties.
It is interesting to note the hype following the introduction of magic angle twists of bilayer structures with graphene and other two-dimensional materials, see Fig.~\ref{fig:bilayer}.
For small twist angles the moir\'{e} patterns lead to large effective unit cells with easily more than \(50-100\) sites and therefore corresponding bands in the band structure.
For fine-tuned magic angles one or a few bands turn almost flat, potentially resulting in superconductivity below one Kelvin.
While this is certainly an important and interesting phenomenon, it is far from the rigor of the flat band physics discussed here and its realization with highly flexible light based artificial material platforms.
It is worth pointing to attempts connecting nearly flat bands in twisted bilayers to perfect flat bands~\cite{tarnopolsky2019origin,ha2021macroscopically}. 

%   Figure 2

The third evolution cycle was ripe to start.
It needed systematic and as complete as possible FB generators.
The key was to use the CLS as a starting point rather than an outcome~\cite{flach2014detangling}.
The next step in that direction successfully lead to the first and complete FB generator for one-dimensional two band lattices~\cite{maimaiti2017compact}. 
With that it was clear that flat band models are and can be obtained through fine-tuning~\cite{maimaiti2019universal,maimaiti2021flatband,graf2021designing,hwang2021general}. 
That result opened the door for experimental platforms that can be accurately fine-tuned.
Indeed classical condensed matter platforms are mostly using electronic and magnonic waves propagating in crystals and materials which lack fine-tuning potential, 
with the spatial periods having length scales of the order of Angstrom and were until recently way beyond the limits of direct in-situ observations~\cite{slot2017experimental,drost2017topological}. 
Instead, some of the most important emerging new photonic platforms are exploiting light propagation in structured media, or atomic Bose-Einstein condensates propagating in optical lattices generated by laser light, 
or light and excitons hybridizing into polaritonic dissipative condensates and propagating in structured media while kept afloat with external laser light pumping~\cite{leykam2018artificial,leykam2018perspective}. 
All these platforms are characterized by a high degree of tunability, and by length scales dictated by the wavelength of light, making direct in-situ observations and characterizations possible.
For example, recent demonstrations of photonic moir\'{e} superlattices illustrate the ability to fabricate fine-tuned periodic structures with hundreds of elements per unit cell~\cite{wang2020localization,mao2021magic,nguyen2022magic,du2023moire}.

This review intends to give a broader view on the fine-tuning properties of flat bands, to discuss in more detail a number of recent fine-tuning results as well as outcomes of perturbing flat bands,
and connect to recent or still to be performed experimental studies in the broader area of photonics and light-matter interaction.

%    ADD  
%    experimental realizations of flatband in photonics~\cite{leykam2018perspective,tang2020photonics} -- photonic waveguide arrays~\cite{nakata2012observation,mukherjee2015observation,mukherjee2015observation1,vicencio2015observation,weimann2016transport}, 
%    photonic crystals~\cite{xia2018unconventional}, 
%    exciton-polariton condensates~\cite{masumoto2012exciton,jacqmin2014direct,biondi2015incompressible,klembt2017polariton}

\section{Needles in the haystack}
%What means flat band fine-tuning and generating?
\label{sec:II}

\begin{figure}[h!]
    \centering
    \includegraphics[width=0.925\columnwidth]{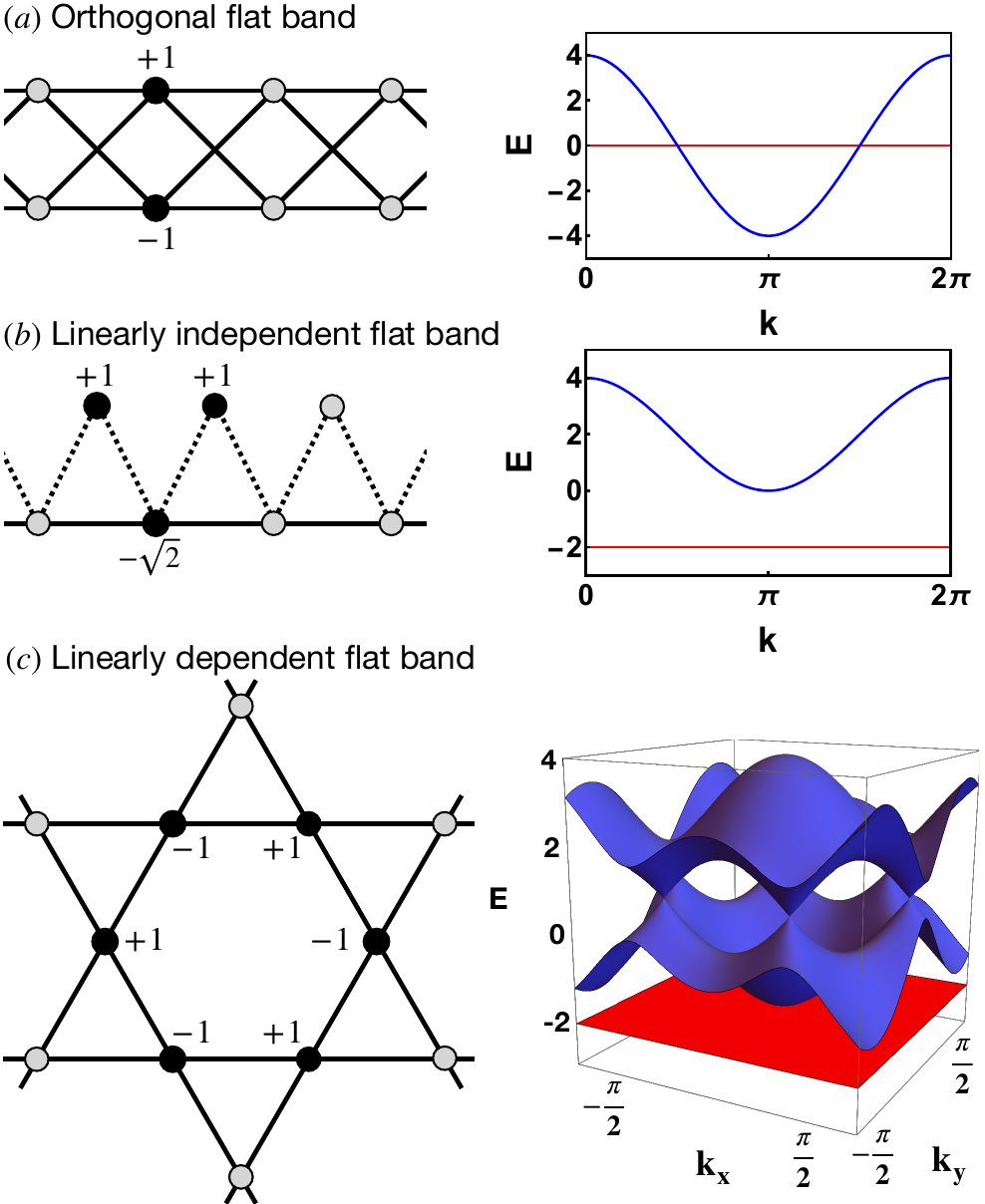}
    \caption{
        (a) A 1d cross-stitch lattice with an orthogonal flat band. 
        (b) A 1d sawtooth lattice with a linearly independent flat band.
        (c) A 2d Kagome lattice with a linearly dependent flat band. 
        For all cases the left plots show the
        lattice structure and location of CLS. 
        Solid circles show the CLS profile with given amplitudes and alternating phases (indicated by the signs).
        Right plots - corresponding band structure for each model (flat band is shown in red color).
    }
    \label{fig:fb-ex}
\end{figure}

Spatially periodic tight-binding networks are models which describe the propagation of linear waves on a spatial lattice.
The wave is described by a finite number \(\nu\) of wave states in a spatial unit cell which is repeated periodically in space. 
Each local wave or mode state is characterized by a complex number - amplitude and phase, similar to an oscillator. 
These wave states are connected through hopping elements with other states and can tunnel (hop), as shown in Fig.~\ref{fig:fb-ex}.
Examples can be e.g. single mode optical waveguides carrying light which tunnels to nearby waveguides through its evanescent tails, among many others.
%electrons hopping in a crystal lattice and resulting in well-known electronic band structures, or 
In real physical space the evolution of these systems is modeled with a time-dependent Schr\"odinger equation \(i\dot{\psi}_{l,\mu} = \sum H_{\{l,\mu\},\{m,\kappa\}} \psi_{m,\kappa}\) with \(H_{\{l,\mu\},\{m,\kappa\}}\) being the matrix element of the Hamiltonian \(\mhsp\). 
Here \(l, m\) label the unit cells, and \(1 \leq \mu,\kappa \leq \nu\) label the state inside one unit cell. 
The onsite energies (diagonal elements) are defined as \(\epsilon_{ l,\mu} =  H_{\{l,\mu\},\{l,\mu\}}\), and the off-diagonal elements describe the hopping, often abbreviated by \(t\) (not to be confused with time). 
Time itself can be the physical time, or the propagation distance of light in the case of optical waveguides. 
Since we consider \(\mhsp\) to be invariant under discrete translations, application of Bloch's theorem and transformation into reciprocal space results in a Hamiltonian \(\mhsp(\vec{k})\). 
The number of allowed wave vectors \(\vec{k}\) equals the number of unit cells, and \(\mhsp(\vec{k})\) is a \(\nu \times \nu\) matrix. 
Its eigenvalues \(E_{\mu}(\vec{k})\) form a Bloch band structure which is periodic in \(k\)-space. 
The normalized eigenvectors are coined the polarization vector field for each of the Bloch bands. 

The networks are classified according to their spatial dimension \(d\), the Bravais lattice structure which is formed by the unit cells, the number of Bloch bands \(\nu\) 
({\it i.e.} the minimal number of sites of the unit cell representation, or likewise the number of sublattices embedded into the Bravais lattice structure) and the hopping range \(m_c\). 
Fixing these classifiers however still leaves the values of the hopping strengths and the onsite energies undefined, yielding a large dimensional manifold of tight-binding networks. 

Without any further restrictions a typical band structure will consist of dispersive bands only.
Viewing such a manifold of in general dispersive band structures as a haystack, the task of finding submanifolds of networks with flat bands in their band structure appears to be similar to the task of finding a needle in the haystack.
We will discuss how this can be done systematically.

The choice of the hopping strengths and the onsite energies becomes crucial to ensure features to a lattice -- {\it e.g.} local and global symmetries, topology, among others. 
Such a procedure of targeting certain features by properly selecting the hoppings and onsite energies, which may either result in a specific lattice model or in sub-families ({\it manifolds}) of lattices, goes by the name of {\it fine-tuning}. 

From a practical viewpoint the fine-tuning of a flat band network therefore is a careful adjusting of the hopping values and onsite energies of a tight-binding network in order to zero the dispersion of (at least) one of its Bloch bands:
\(E_\mu(\vec{k}) = E_\mu(\vec{k}')\) for any choice of the wave vectors. 
%Careful adjusting refers to an experimental approach in a lab, where the hopping strengths of a lattice have to be adjusted one at the time by {\it e.g.} tuning the distance between neighboring waveguides. 
%From a theoretical standpoint instead, fine-tuning of flat band networks means to find the conditions on both hopping and onsite energies so that a lattice possesses (at least) one flat band.
Identifying these fine-tuning conditions is equivalent to obtaining a flat band generator: fixing relevant classifiers as e.g. the desired dimension \(D\), number of bands \(\nu\), and hopping range \(m_c\) of the lattice, {\it generate} the manifold of networks \(\mhsp\) which possess (at least) one flat band.
If the flat band generator approach results in reliable predictions, they can be applied in the lab to fine-tune samples in order to obtain flat bands.

Making a network to possess certain global symmetries can be viewed on one hand as a special way of finetuning, and on the other hand as a way to ensure the existence of flat bands and CLSs without the need of expliticely proving their existence.
%AA: what about the non-Hermitian \(\mathcal{PT}-symmetry\)?
%PSF: T does not relate to CLS and FBs.
Two known cases - chiral~\cite{ramachandran2017chiral} and anti-parity-time~\cite{mallick2022antipt} symmetries - are doing the job.
Let us consider the chiral case, and use the 2D Lieb lattice in Fig.~\ref{fig:lieb}(a) as an example.
For that we need bipartite networks with all sites separated into sublattices A and B (red and blue sites in Fig.~\ref{fig:lieb}), such that sites from sublattice A are connected only to sites from sublattice B and vice versa, without any additional onsite energies.
Eigenstates of such a network come in pairs with energy \(\pm E\).
We further request to have more sublattice A sites in a unit cell than sublattice B sites, making the A sublattice a so-called majority sublattice.
In that case it follows~\cite{ramachandran2017chiral} that there must be a macroscopic number (at least equal to the number of unit cells) of eigenstates with chiral energy \(E=0\), which therefore form a chiral flat band.
It further follows that the eigenvectors are nonzero entirely on the majority sublattice, and that they can be cast into the form of CLSs.
Adding hoppings which connect the minority sublattice sites and onsite energies on these sites will not affect the chiral flat band and its CLSs.
Making all hoppings as well as the minority site energies random will not destroy the CLSs either.
Shifting the majority onsite energies equally will also keep the CLSs and only shift the flat band energy~\cite{calugaru2022general}.

%   Figure 3

\section{Flat band types}
\label{sec:III}

\begin{figure}[h!]
    \centering
    \includegraphics[width=0.925\columnwidth]{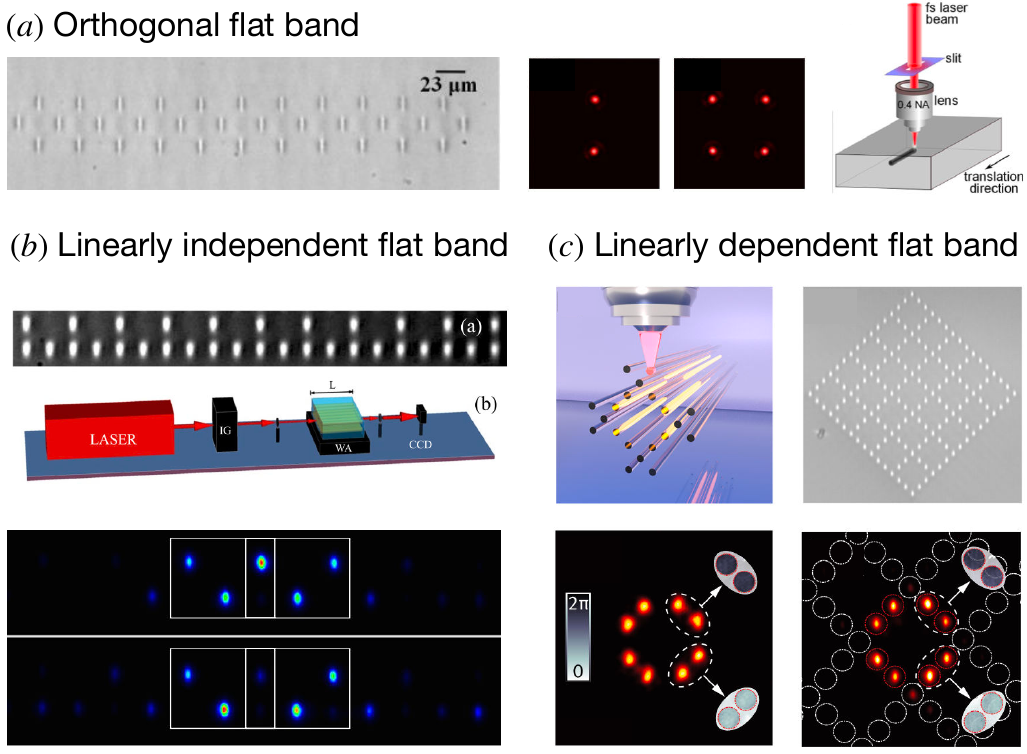}
    \caption{
        Experimental realizations of samples of the three types of flat bands.
        (a) A 1d rhombic lattice with an orthogonal flat band, adapted from Ref.~\cite{mukherjee2015observation1}. 
        (b) A 1d Stub lattice with a linearly independent flat band, adapted from Ref.~\cite{real2017flat}. 
        (c) A 2d decorated Lieb lattice with a linearly dependent flat band, adapted from Ref.~\cite{hanafi2022localized}.
    }
    \label{fig:fb_experiments}
\end{figure}

For networks with finite hopping range \(m_c<\infty\) and a finite number of bands \(\nu < \infty\), a flat band results in the existence of a macroscopically degenerate set of compact localized eigenstates~\cite{read2017compactly}. 
Enforcing the existence of compact localized states in tight-binding networks yields an effective and systematic method for generating flat band. 
% in any \(D\), \(\nu\). 
This approach calls for additional classifiers for flat bands in addition to the above-mentioned ones \(D\), \(\nu\), \(m_c\) -- namely the {\it size} \(U\) of a CLS ({\it i.e.} the number of occupied unit cells) and for \(D\geq 2\) the {\it shape} ({\it i.e.} the arrangement of such \(U\) unit cells). 
% As discussed later in Sec.~\ref{sec:IV_1}, the choice of such classifiers has strong impact on the features of both the set of CLSs and the overall band structure.  
% The amplitudes of the encoded CLS profile become the knobs of the generator scheme, as tuning such entrees yields different flat band networks. 
% The resulting parametrization scheme of flat band networks is exhaustive and applicable in any dimension \(D\) and for any finite number of bands \(\nu\). It further provides a useful tool to study and control the impact of perturbations -- as it will be discussed in Sec.~\ref{sec:IV_2} and in Sec.~\ref{sec:V} concerning onsite disorder and interactions respectively.   
%For a fixed set of parameters \(D, \nu, m_c\) generic randomly chosen translationally invariant tight-binding networks have only dispersive Bloch bands.
%Fine-tuning of the hoppings and onsite energies is required to flatten one or several of the bands.
%Flat band lattices with finite-range hopping feature compact localised states.
%While the CLS are defined in real space, they are also directly imprinted in the corresponding polarisation vectors in the Bloch space and can be read off directly from those.
%We also define a {\it local unitary transformation} -- a unitary transformation, that only involves a strictly finite number of sites, that is used below.
%
%
%
Flat bands can be sorted into different classes of varying complexity, that largely define their response to perturbations.
These classes are defined by the properties of the CLS set of a flat band: their {\it orthogonality} and {\it completeness}, i.e. {\it linear independence}.
Further the classes differ by the result of a projection of an arbitrary state on the FB state space, i.e. by the so-called real space projector.
Finally FBs from different classes respond differently to the action of local unitaries, i.e. unitary transformations of the unit cell (and subsequent local unitaries after redefining the unit cell).
The simplest class consists of {\it orthogonal flat bands}~\cite{flach2014detangling}, whose set of CLS is complete and orthonormal, for example the flat band of the cross-stitch lattice (see Fig.~\ref{fig:fb-ex}(a)).
Real space projectors onto the flat band Hilbert space are compact in this case.
Local unitaries can completely detangle the FB and its CLSs from the rest of the network, i.e. they can result in a diagonalization of the FB part using strictly local transformations, without the need to transform into \(k\)-space~\cite{flach2014detangling}.
%or the \(E=0\) flat band of the diamond chain~\cite{} (see Fig.~\ref{fig:fb-ex}(b)).
Next is the class of {\it linearly independent flat bands} with a complete and linearly independent but non-orthogonal set of CLS, like the sawtooth lattice~\cite{huber2010bose} (see Fig.~\ref{fig:fb-ex}(b)).
For this class, there is no {\it local} unitary transformation that can make the CLS set orthogonal.
%Locality of the transformation is important, since there are nonlocal transformations achieving orthogonality.
%The obvious example is the Fourier transform into the Bloch space, which yields an orthogonal set of eigenstates, that are no longer local in real space. 
The real space projector for linearly independent flat bands is decaying exponentially in space.
A detangling procedure is not applicable for this class.
%with an exponent which is directly related to the distance of such a gapped flat band from any other band in the spectrum.
Finally, there are {\it linearly dependent flat bands} whose CLS sets are neither orthogonal nor complete, but linearly dependent like that of the Kagome lattice~\cite{lieb1989two} (see Fig.~\ref{fig:fb-ex}(c)).
Linearly dependent flat bands exist only in dimension \(d\geq 2\), and are touching at least one dispersive band at a special \(k\) point.
Their real space projector is decaying algebraically as \(1/r^d\)~\cite{chalker2010anderson,yan2023classification,yan2023classification2}.
Such bands are also referred to as critical or singular~\cite{rhim2019classification,rhim2021singular}, since their polarisation vector in the Bloch space vanishes in some point in the Brillouin zone, indicating a band touching.
Conversely, all the flat bands with polarisation vectors that are nonzero everywhere in the Brillouin zone, e.g. with complete sets of CLS~\cite{rhim2019classification}, are called non-singular.
Experimental photonic realizations of the three types of flat bands were reported in Refs.~\cite{mukherjee2015observation1,real2017flat,hanafi2022localized} and are shown in Fig.~\ref{fig:fb_experiments}.

%   Figure 4

\begin{figure}[h!]
    \centering
    \includegraphics[width=0.925\columnwidth]{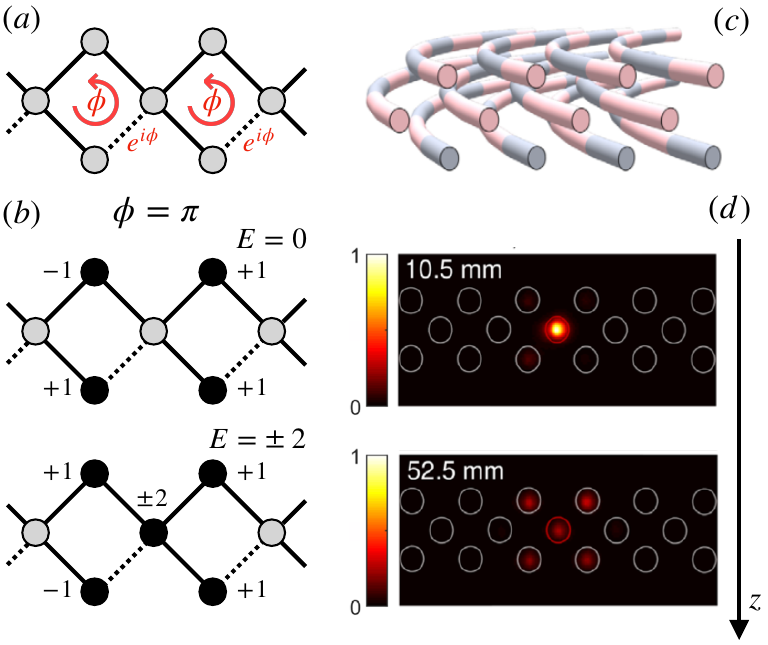}
    \caption{
        (a) Diamond chain profile with magnetic flux \(\phi\) (dashed complex hoppings).
        (b) CLS profiles for the all bands flat with flux value \(\phi=\pi\) and three flat bands at \(E=0\) and \(E=\pm 2\). 
        (c) Simplified sketch illustrating the waveguide paths of a photonic diamond chain with \(\phi=\pi\) flux per plaquette realized by circularly curving the lattice, adapted from Ref.~\cite{mukherjee2018experimental} 
        (d) Experimentally measured output intensity distributions at two different propagation distances, adapted from Ref.~\cite{mukherjee2018experimental}. 
    }
    \label{fig:fb_abf}
\end{figure}

A particular type of networks host a finite number of bands with all bands flat (ABF).
Flat bands from ABF lattices fall into the first class of orthogonal ones \cite{danieli2021nonlinear,sathe2023topological}.
Since all bands lack dispersion, there is no transport in an ABF network. Consequently any local initial state will not be able to propagate beyond a certain finite distance. 
Early realizations were theoretically obtained by Vidal et al~\cite{vidal1998aharonov,vidal2000interaction} through adding a magnetic flux to a bipartite flat band lattice with the flat band at energy \(E=0\) being protected by chirality~\cite{ramachandran2017chiral}.
The flux only modifies the hoppings, preserves the chiral symmetry, and keeps the \(E=0\) band flat.
An example is shown in Fig.~\ref{fig:fb_abf}(a) for the diamond chain.
At the magic flux value \(\phi=\pi\) all other bands turn flat as well, resulting in an ABF lattice with flat band energies \(E=0,\pm2\).
The CLS profiles are shown in Fig.~\ref{fig:fb_abf}(b).  
Experimental photonic realizations of fluxes and ABF lattices were obtained by Mukherjee et al~\cite{mukherjee2018experimental} with the help of circularly curved laser-imprinted photonic waveguides (Fig.~\ref{fig:fb_abf}(c)).
Careful tuning to the magic flux value \(\phi=\pi\) allowed to observe absence of dispersion and complete localization and trapping of a locally injected light excitation as shown in Fig.~\ref{fig:fb_abf}(d).
Other experimental realizations of ABF lattices in photonic lattices include~\cite{kremer2020square,Jorg2020artificial,caceres2022controlled}. 

%As discussed later in Sec.~\ref{sec:IV_1}, t
The choice of flat band classifiers has a strong impact on the features of both the set of CLSs and the overall band structure.  
The amplitudes of the encoded CLS profile become the knobs of generator schemes, as tuning such entries yields different flat band networks. 
The resulting parametrization scheme of flat band networks is exhaustive and applicable in any dimension \(D\) and for any finite number of bands \(\nu\). It further provides a useful tool to study and control the impact of perturbations.
%-- as it will be discussed in Sec.~\ref{sec:IV_2} and in Sec.~\ref{sec:V} concerning onsite disorder and interactions respectively.   

% We can conclude that the needle in the haystack - the flatband manifold - has a surprising internal fine structure. Properly identifying this manifold substructure promises to observe novel transport regimes which emerge from perturbations of different flatband submanifolds. Even if the perturbation is the same, the outcome can be expected to differ qualitatively depending on the submanifold it is applied to.

%   Figure 5

%\begin{enumerate}
%    \item 
%    in general, translationally invariant lattices posses non-zero Bloch dispersion relations

%    \item
%    flatband networks with Hamiltonian \(\mhsp\) result from hopping fine-tuning aimed to vanish the dispersion of at least one Bloch band
    
%    \item 
%    different classes of flatbands (orthogonal, non-orthogonal \& linearly independent, non-orthogonal \& linearly dependent), correspondent singular/non-singular polarization vector fields
%\end{enumerate}

\section{Tuning and detuning}
\label{sec:IV}

The needle in the haystack - the flat band manifold - has a surprising internal fine structure.
Properly identifying this manifold substructure promises to observe novel transport regimes which emerge from perturbations of different flat band submanifolds.
Even if the perturbation is the same, the outcome can be expected to differ qualitatively depending on the submanifold it is applied to.

\subsection{Tuning}
\label{sec:IV_1}

To achieve fine tuning - or simply tuning - is synonymous to finding a generator procedure which has flat band network models on output.
What sets different generators apart is not only the algorithm itself, but also the things which need to be specified on input.

\begin{figure}[h!]
    \centering
    \includegraphics[width=0.95\columnwidth]{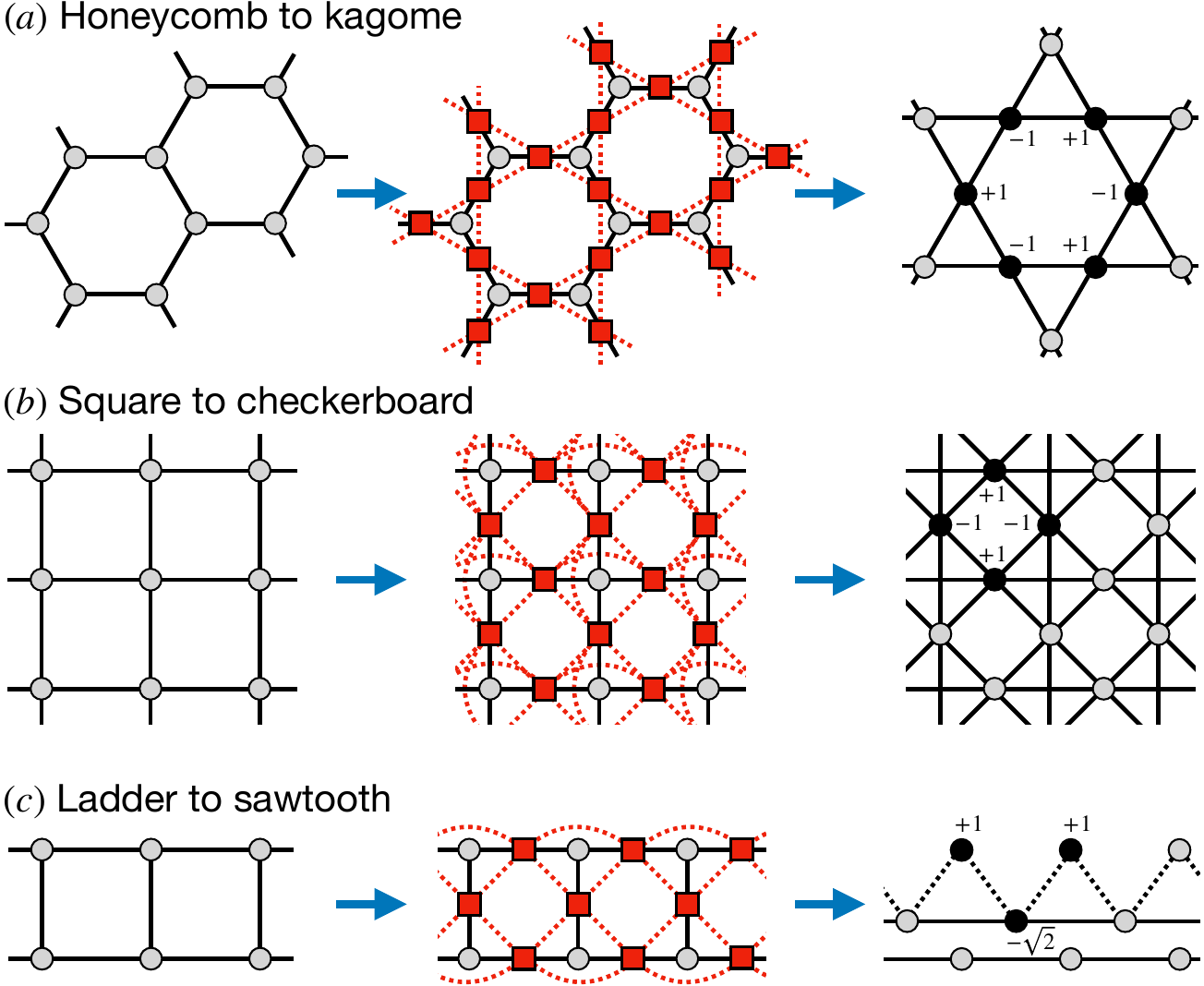}
    \caption{
        Line graph constructions of the kagome lattice from the honeycomb lattice (a), of the checkerboard lattice from the square lattice (b), and the sawtooth chain from the two legs ladder (c).
    }
    \label{fig:fb-mielke}
\end{figure}

An early documented attempt is the line graph generator of Mielke~\cite{mielke1991ferromagnetism} designed to tackle flat band ferromagnetism in Hubbard models. 
On input one chooses a tight binding network with some dispersive band structure which does not contain a flat band.
One then constructs the line graph of this network.
It is obtained by replacing bonds by new sites, and new bonds (all of the same amplitude) between all new sites whose old bonds shared a joint old site.
The line graph is a new tight binding network which possesses one flat band as the lowest energy band in the band structure (or highest, depending on the new bond strength sign)~\cite{mielke1991ferromagnetism}.
It was mainly studied and applied in two dimensions.
Examples are the line graph of the honeycomb lattice which is the kagome lattice (Fig.~\ref{fig:fb-mielke}(a)), and the square lattice maps into the checkerboard lattice (Fig.~\ref{fig:fb-mielke}(b)).
But it works also in one dimension, where e.g. a simple two leg ladder will map onto its line graph - the sawtooth chain (Fig.~\ref{fig:fb-mielke}(c)).
This line graph generator does not contain continuous free parameters, thus it generates a countable set of flat band networks only.
Further the flat band is always located at the extreme of the band structure.

\begin{figure}[h!]
    \centering
    \includegraphics[width=0.9\columnwidth]{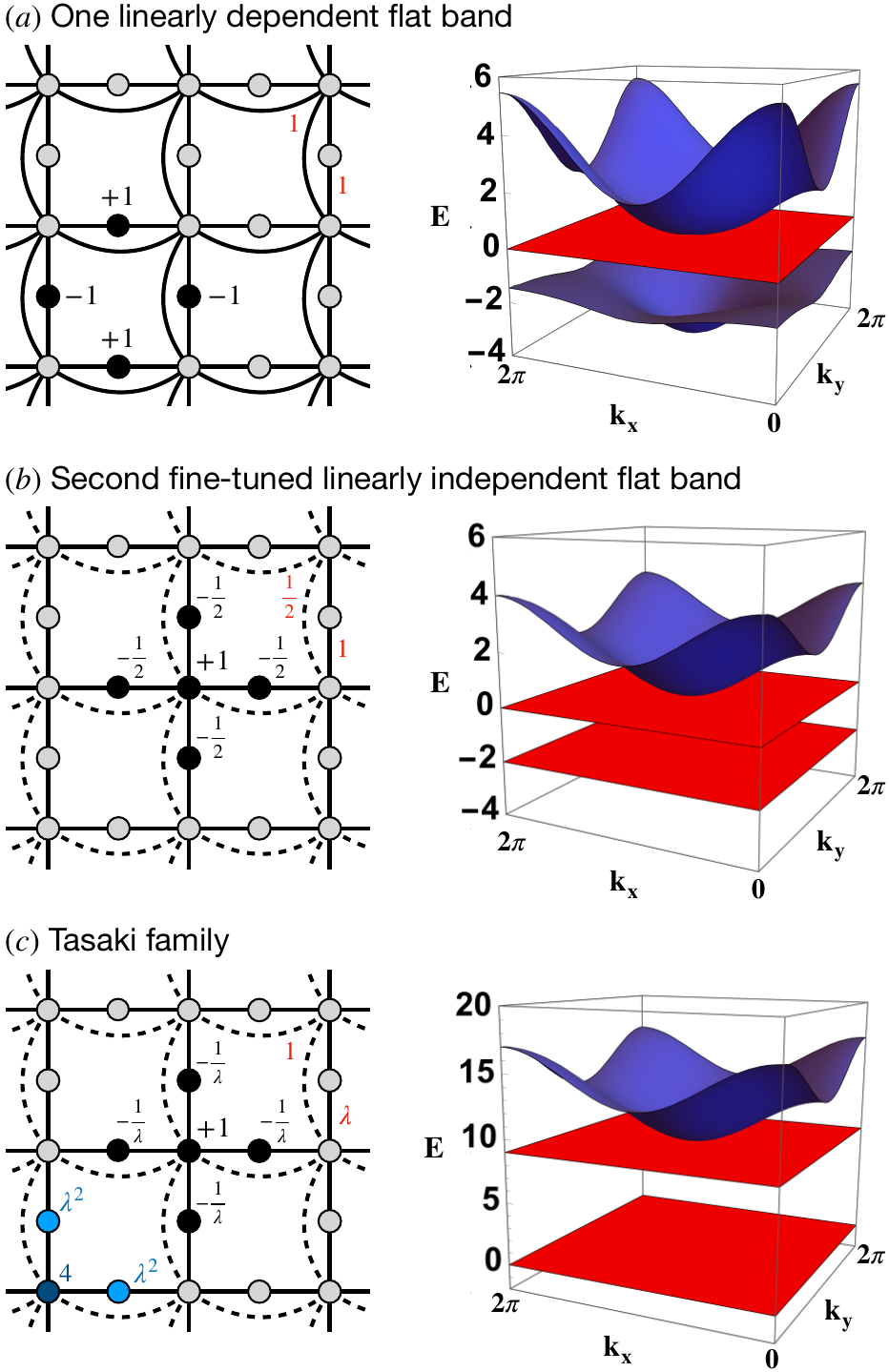}
    \caption{
        (a) Decorated 2d Lieb (left) with uniform hopping strengths. Solid circles show the CLS profile with given amplitudes and alternating phases (indicated by the signs). 
        Band structure (right) with one flat band (red) band touching the upper dispersive band (blue) and gapped away from the lower dispersive band (blue). 
        (b) Same as (a) with fine-tuned next nearest neighbor hopping to \(1/2\).
        Solid circles show the CLS profile corresponding to the second flat band at \(E=-2\).  
        (c) Tasaki family (left) with parametric hopping \(\lambda\) (solid line) and onsite energies (blue colored circles). 
        Band structure (right) with two flat bands (red) and a dispersive band (blue) for \(\lambda=3\). 
    }
    \label{fig:fb-tasaki}
\end{figure}

Another early achievement was due to Tasaki~\cite{tasaki1992ferromagnetism}.
While mainly focusing on flat band ferromagnetism aspects due to many-body interactions, this work studied the consequences of decorating the known 2D Lieb flat band lattice (Fig.~\ref{fig:lieb}) with next to nearest neighbour hoppings between minority sublattice sites as shown in Fig.~\ref{fig:fb-tasaki}(a).
As discussed above, this modification destroys the chiral symmetry, yet preserves the flat band at \(E=0\) with its unchanged CLS.
Since the chirality is broken, we can expect that the two dispersive bands open a gap between each other, and cease to be symmetry related copies of each other.
However the CLS of the flat band is unchanged - therefore the CLS set must still be linearly dependent!
From that we conclude that the flat band at \(E=0\) must still touch one of the dispersive bands, as confirmed in Fig.~\ref{fig:fb-tasaki}(a).
A further surprise comes when choosing the additional hoppings to be half of the original Lieb hoppings.
In that case lower dispersive band turns into a second flat band.
It is supported by a new CLS in Fig.~\ref{fig:fb-tasaki}(b).
Tasaki also added and finetuned onsite energies and managed to obtain a flat band network for which the new ground state flat band is supported by tunable CLS sets as shown in Fig.~\ref{fig:fb-tasaki}(c).
The main conclusion then is that flat band networks have to be finetuned on the one hand with respect to their network parameters, but form continuous manifolds on the other hand with some tuning and navigating manifold parameters which smoothly change the CLS details.
This insight culminates in the discovery of flat band generators which are based on the properties of tunable CLS sets to be supported.

\begin{figure}[h!]
    \centering
    \includegraphics[width=0.9\columnwidth]{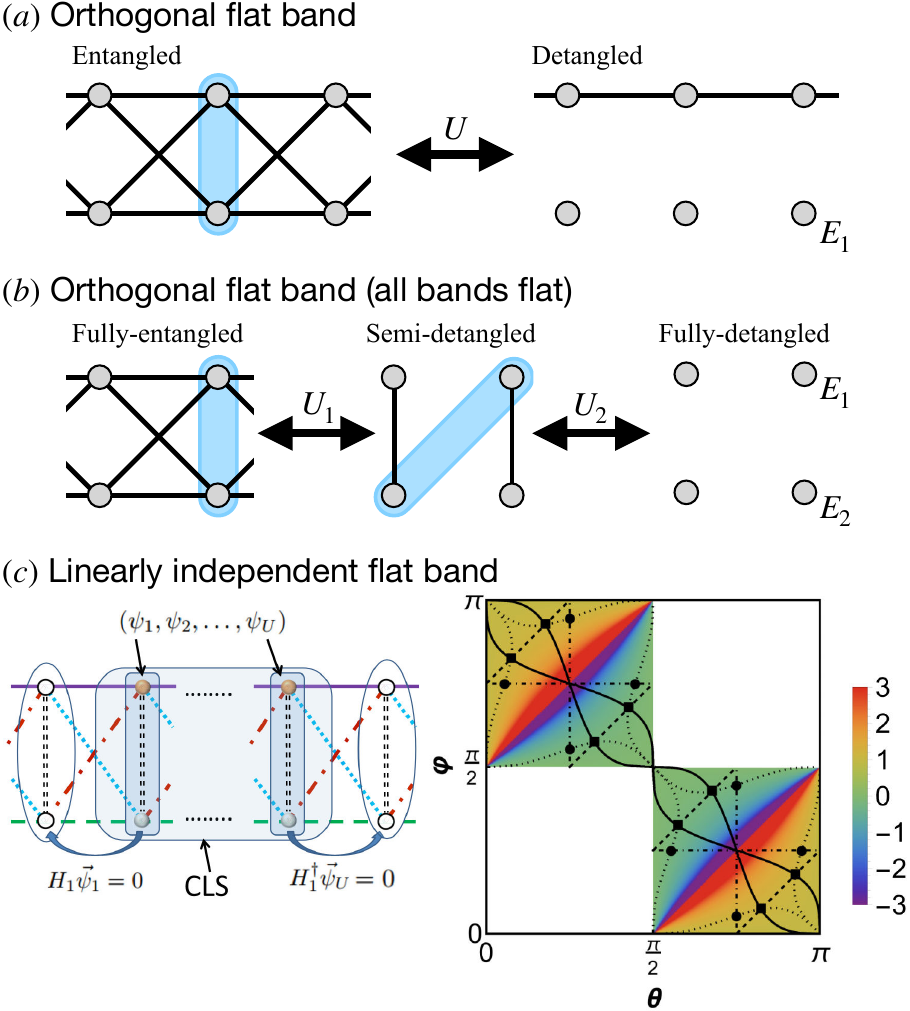}
    \caption{
        (a) Entangling procedure for lattices supporting one dispersive band and one flat band with orthogonal CLS via local a  unitary transformation within each unit cells (blue shaded area). 
        (b) Same as (a) for all-band-flat network. 
        (c) Left: schematics of the compact localized state for nearest-neighboring lattice. 
        Right: flat band energy for two bands lattice as function of the parameterizing angles. 
        The colored squares host FB networks, while the white ones do not (for nearest-neighboring hopping).
        The solid and dashed lines indicate diverse flat band sub-manifolds.
        Figure adapted from Ref.~\cite{maimaiti2017compact}. 
    }
    \label{fig:fb-entangling}
\end{figure}

CLS based generators for orthogonal flat bands have been reported in Ref.~\cite{flach2014detangling}.
They harvest on the existence of local unitary operations (one per each unit cell) which completely detangle the CLSs into uncoupled single sites in the new basis.
Inverting this detangling procedure generates tunable families of flat band networks.
One therefore starts with a choice of some dispersive tight binding network, adds one or more decoupled sites (at possibly one or more different onsite energies) to each unit cell, and then applies the local unitaries.
In Fig.~\ref{fig:fb-entangling}(a) we schematically show this detangling-entangling procedure for a 1D two bands problem with one flat band. 
The free parameters are the angles of the local unitary transformation.
In addition one can redefine the unit cell and perform another tunable unitary procedure.
The resulting CLS will also continuously depend on the tuning parameters, and increase in size the more redefined unit cells are used.
Still both the CLS set stays orthonormal, and the flat band energies are kept unchanged.
The procedure works in any lattice dimension, for any lattice to start with.
To generate ABF networks one has to only choose a Bravais lattice, the number of flat bands and their flat band energies.
Each Bravais lattice point is then initially containing only uncoupled sites at their own onsite energies, one per flat band.
Then one performs at least two sets of local unitary operations (in 1D, more are required in higher dimensions) with different unit cell choices to arrive at a fully connected network, yet with all bands flat~\cite{danieli2021nonlinear,danieli2021quantum} 
as shown in Fig.~\ref{fig:fb-entangling}(b).
Early studies of Vidal et al~\cite{vidal1998aharonov,vidal2000interaction} as well as later ones by Brosco et al~\cite{brosco2021twoflux} added magnetic fluxes to chiral flat band networks and observed that at particular values of the flux all bands turn flat, which was coined Aharonov-Bohm caging.
It follows that there is no need to introduce first a magnetic flux and then to finetune this flux value to arrive at ABF networks.
Instead one can simply follow the above local unitary approach to arrive at the most complete ABF generator.

%   Figure 8

CLS based generators for non-orthogonal but in general linearly independent flat bands are the most involved ones.
To succeed, one needs to first choose a particular CLS and a particular network class with still free hopping strength parameters -- as schematically represented in the left panel of Fig.~\ref{fig:fb-entangling}(c).
Second, one formulates hopping strength conditions for the destructive interference at the surface of the CLS.
Finally, one needs to satisfy the hopping strength conditions on a finite network manifold of the size of the CLS which supports this CLS as an eigenstate. 
The corresponding algebraic equations which need to be solved are usually of nonlinear nature, which can complicate the task.
For the simplest case of dimension \(d=1\), nearest unit cell hopping only, and two bands with one of them flat, the task can be accomplished in closed form analytically as shown by Maimaiti et al~\cite{maimaiti2017compact}. 
Interestingly the entire family can be mapped onto 1D sawtooth networks Fig.~\ref{fig:fb-ex}(b) with tunable hoppings and onsite energies. 
Two nontrivial free angles remain which navigate through the flat band family as shown in Fig.~\ref{fig:fb-entangling}(c) (colored areas). 
In particular, the solid and dashed lines indicate subfamilies of flat band networks with symmetry relations between hoppings and potentials.
The upshot then is that indeed tunable CLS sets lead to tunable flat band networks - the needle in the haystack has an enormous internal structure.
Maimaiti et al extended the generator method to 1D networks with more bands~\cite{maimaiti2019universal}, as well as to 2D networks~\cite{maimaiti2021flatband} confirming the above conclusions.
The algebraic complexity quickly grows and calls for special ways to solve the resulting equations.
Nonorthogonal flat band generators can be always combined with orthogonal flat band generators, either in order to complexify the resulting network structure even more, or in order to perform a partial detangling of a macroscopic fraction of CLSs for non-orthogonal flat bands~\cite{maimaiti2017compact}.

In order to generate linearly dependent flat band networks, all one needs is to choose CLS sets which are linearly dependent, and then apply the above generator scheme for non-orthogonal flat bands.
For 1D networks it was shown by Maimaiti et al that a linear dependent CLS set can be reduced to a set with smaller CLSs, and in the case of band touching a dispersive band or even crossing it to an orthonormal CLS set~\cite{maimaiti2017compact,maimaiti2019universal}.
In higher dimensions linearly dependent flat bands lead to band touching which can be separated additionally into linear touching (with two dispersive bands e.g. Fig.~\ref{fig:lieb}, or quadratic touching (with only one dispersive band, e.g. Fig.~\ref{fig:fb-ex}(c) and Fig.~\ref{fig:fb-tasaki}. 
Recent studies of Graf et al~\cite{graf2021designing} and Hwang et al~\cite{hwang2021general} use the transformation of CLSs into Bloch representation to formulate linear dependence conditions in reciprocal space and to come up with a final generator of linearly dependent flat band networks.

The above generators can also be extended to the case of non-Hermitian Hamiltonians with flat bands~\cite{maimaiti2021nonhermitian}, relevant in photonics.
Non-Hermitian flat bands keep some of the above listed properties of the Hermitian flat bands, e.g. formation of continuous manifolds, but further studies are required.
Also a number of further more specialized flat band generators have been reported, among them the origami rules~\cite{dias2015origami}, repetitions of mini-arrays~\cite{morales2016simple}, Moir\'e engineering~\cite{zhou2023generation}, local symmetry partitioning~\cite{rontgen2018compact}, and path-exchange~\cite{bae2023isolated}.

We close this section with discussing the case of Wannier-Stark (WS) flat bands~\cite{mallick2021wannier}.
One starts with a tight binding network on a Bravais lattice in dimension \(d\geq 2\), e.g. the simple 2D square lattice with nearest neighbour hopping.
Its band structure consists out of one band with energy depending on the two components of the Bloch momentum vector.
Next one applies a dc field along some direction, which induces onsite energies growing linearly along this field direction.
We can consider the field direction as a tunable parameter.
Translational invariance is broken along the field direction in general.
However the translational invariance perpendicular to the field direction can be preserved (or fine-tuned) for certain rational field directions - actually for an infinite amount of them.
These cases lead to an infinite equidistant ladder of WS bands as functions of some perpendicular Bloch momentum vector of dimension \((d-1)\).
However each eigenstate is still \(d\)-dimensional, even though its amplitudes along the field direction decay superexponentially in both directions.
As shown by Mallick et al in Ref.~\cite{mallick2021wannier}, almost all rational field directions (except for a few) will result in strictly flat WS bands, such that the WS band structure will consist of an infinite ladder of flat bands only.
Transport is then completely suppressed.
Yet the WS flat bands do not support CLSs, instead the wave function localizes at best superexponentially in all directions.
This outcome is not in contradiction with above statements that ABF flat bands are orthonormal and support CLSs, since one condition for that is a finite number of bands.
For WS flat bands the number of bands in the WS ladder is infinite instead.
Finally we note that WS flat bands can be mapped onto Floquet systems in \(d-1\) dimensions.

\subsection{Detuning}
\label{sec:IV_2}

\begin{figure}[h!]
    \centering
    \includegraphics[width=0.9\columnwidth]{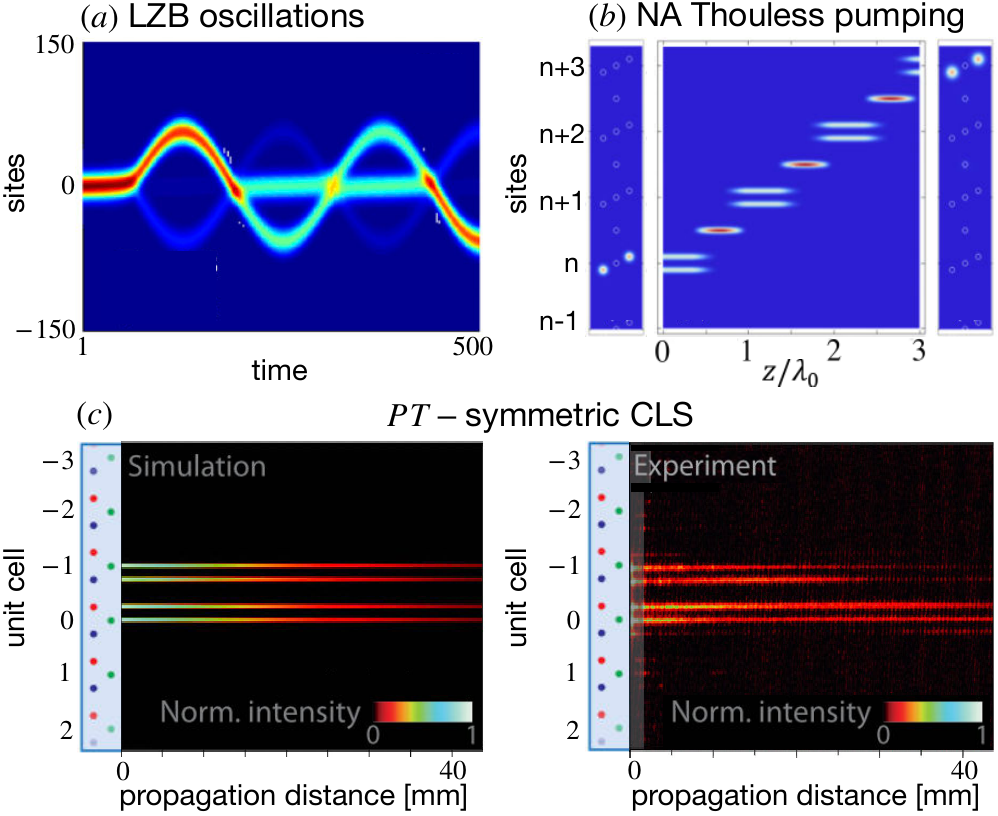}
    \caption{
        (a) Landau-Zener-Bloch oscillations of an excited CLS in a diamond chain with non-zero perpendicular and longitudinal dc electric and magnetic fields.
        Adapted from Ref.~\cite{khomeriki2016landau}. 
        (b) Non-abelian Thouless pumping of a CLS in a 1D stub-like chain via periodic modulation of the hopping terms of period \(lambda_0\).
        Adapted from Ref.~\cite{Brosco2021nonabelian}.
        (c) Diffraction-free propagation of an excited CLS in a \(\mathcal{PT}\)-symmetric saw-tooth-like photonic flat band lattice.
        The numerical simulation is shown in the left panel, and the experimental observation is shown in the right panel.
        Adapted from Ref.~\cite{tobias2019experimental}.  
    }
    \label{fig:FB_detuning}
\end{figure}

A flat band is fragile against perturbations, raising the question of how a flat band system responds to perturbations.
Various types of perturbations of flat band Hamiltonians which can be experimentally implemented with photonics platforms have been considered over the years — {\it e.g.} dc fields, periodic driving, dissipative potentials.
It emerged that, when properly arranged, these terms can result in unusual phases and phenomena. 
For instance, a properly tuned interplay between magnetic and two dc fields (perpendicular and longitudinal) yields Landau-Zener-Bloch oscillations in a 1D diamond chain ~\cite{khomeriki2016landau,xia2021band}, as shown in Fig.~\ref{fig:FB_detuning}(a). 
Similarly, Bloch oscillations have been ignited in 1D Lieb lattices via non-hermitian driving~\cite{xia2021higher}.
Fine-tuned driving protocols of the hopping terms of flat band lattices can also result in displacing CLS throughout the network while retaining their spatial profile. 
This for instance resulted into the storage and transfer of CLS in the 2D Lieb~\cite{rontgen2019quantum} and non-abelian Thouless pumping of for instance CLS in a 1D stub-like chain~\cite{Brosco2021nonabelian} -- with the latter one shown in Fig.~\ref{fig:FB_detuning}(b). 
Non-Hermitian~\cite{leykam2017flat,danieli2020casting} and \(\mathcal{PT}\)-symmetric~\cite{ge2015parity,molina2015flat} perturbations have been studied in flat band networks. 
In Ref.~\cite{tobias2019experimental}, exact CLS in \(\mathcal{PT}\)-symmetric saw-tooth-like photonic flat band lattice have been theoretically predicted and experimentally verified (Fig.~\ref{fig:FB_detuning}(c)). 
It was also shown that properly tuned gain/loss terms can induce defect states~\cite{qi2018defect} and stable laser emissions~\cite{longhi2019photonic}.

%   Figure 9

\begin{figure}[h!]
    \centering
    \includegraphics[width=0.9\columnwidth]{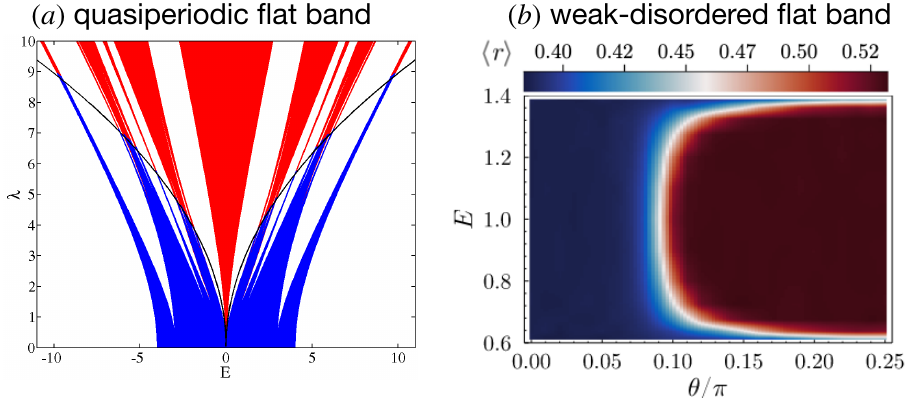}
    \caption{
        Fine-tuned metal-insulator transition disordered one-dimensional \(\nu=2\) flat band networks.
        (a) Cross-stitch lattice with anti-symmetric quasiperiodic perturbation of strength \(\lambda\). 
        The analytic mobility edge (black line) separates extended (blue) and localized (red) phases.
        Adapted from Ref.~\cite{bodyfelt2014flatbands}. 
        (b) Energy-resolved exponent average ratio of adjacent gaps \(\langle r\rangle\) at weak disorder limit  as a function of the ABF manifold angle \(\theta\).
        Adapted from Ref.~\cite{cadez2021metal}. 
    }
    \label{fig:fb-mit}
\end{figure}

An important class of perturbations is {\it disorder}, where the response proved to be especially rich, depending both on the type of the flat band and the type of disorder. 
The impact of uncorrelated disorder has been broadly studied
~\cite{bae2023isolated,leykam2013flat,leykam2017localization,bilitewski2018disordered,Mao2020disorder,Liu2020localization,Liu2021localization}. 
In particular, the weak disorder regime demonstrated unconventional scaling of the localization length in 1D lattices hosting both orthogonal~\cite{leykam2013flat,flach2014detangling} and non-orthogonal flat bands~\cite{leykam2017localization}. 
Anomalous scalings have been also observed in higher dimensional lattices — {\it e.g} in 2D Lieb~\cite{Mao2020disorder} as well as in 3D Lieb lattices~\cite{Liu2020localization,Liu2021localization} due to uncorrelated disorder. 
%
%An important class of perturbation is {\it disorder}, where the response proved to be especially rich, depending both on the type of the flatband and the type of disorder.
%The weakly disordered orthogonal flatband case demonstrated unconventional scaling of localization length with uncorrelated disorder~\cite{leykam2017localization}.
The introduction of correlations, e.g. quasiperiodicity, can introduce mobility edges~\cite{bodyfelt2014flatbands,danieli2015flatband} as shown in Fig.~\ref{fig:fb-mit}(a).
Correlations in the disorder might also induce the inverse Anderson transition~\cite{liu2022unconventional}.
Flat bands with a band touching, non-orthonormal and incomplete set of CLS, can induce critical states at the FB energy in presence of the weak uncorrelated onsite disorder~\cite{chalker2010anderson}.
In some scenarios flat bands turned out to be immune to disorder due to a symmetry protection~\cite{ramachandran2017chiral,liu2022unconventional}.
All-bands-flat Hamiltonians are especially sensitive to different types of disorder~\cite{gligoric2020influence}.
In particular, they can exhibit a metal-insulator transition for weak/infinitesimal disorder driven by the position of the Hamiltonian on the ABF manifold~\cite{cadez2021metal} as illustrated in Fig.~\ref{fig:fb-mit}(b).
As a consequence, the ABF networks with a metallic phase at weak disorder undergo a re-entrant localization transition at finite disorder strengths~\cite{goda2006inverse,longhi2021inverse,li2022aharonov}.
Quasiperiodic perturbations of all-bands-flat networks lead to the emergence of critical phases and fractal edges separating critical and localised eigenstates in the spectrum, in analogy to mobility edges~\cite{lee2023critical,ahmed2022flat}.
The weak disorder results can be rationalised in the framework of projected models, where one considers a projection onto the flat band resulting in an effective single band model with finite and correlated disorder.

%   Figure 10

\subsection{Applications in photonics}
\label{sec:IV_3}

\begin{figure}[h!]
    \centering
    \includegraphics[width=\columnwidth]{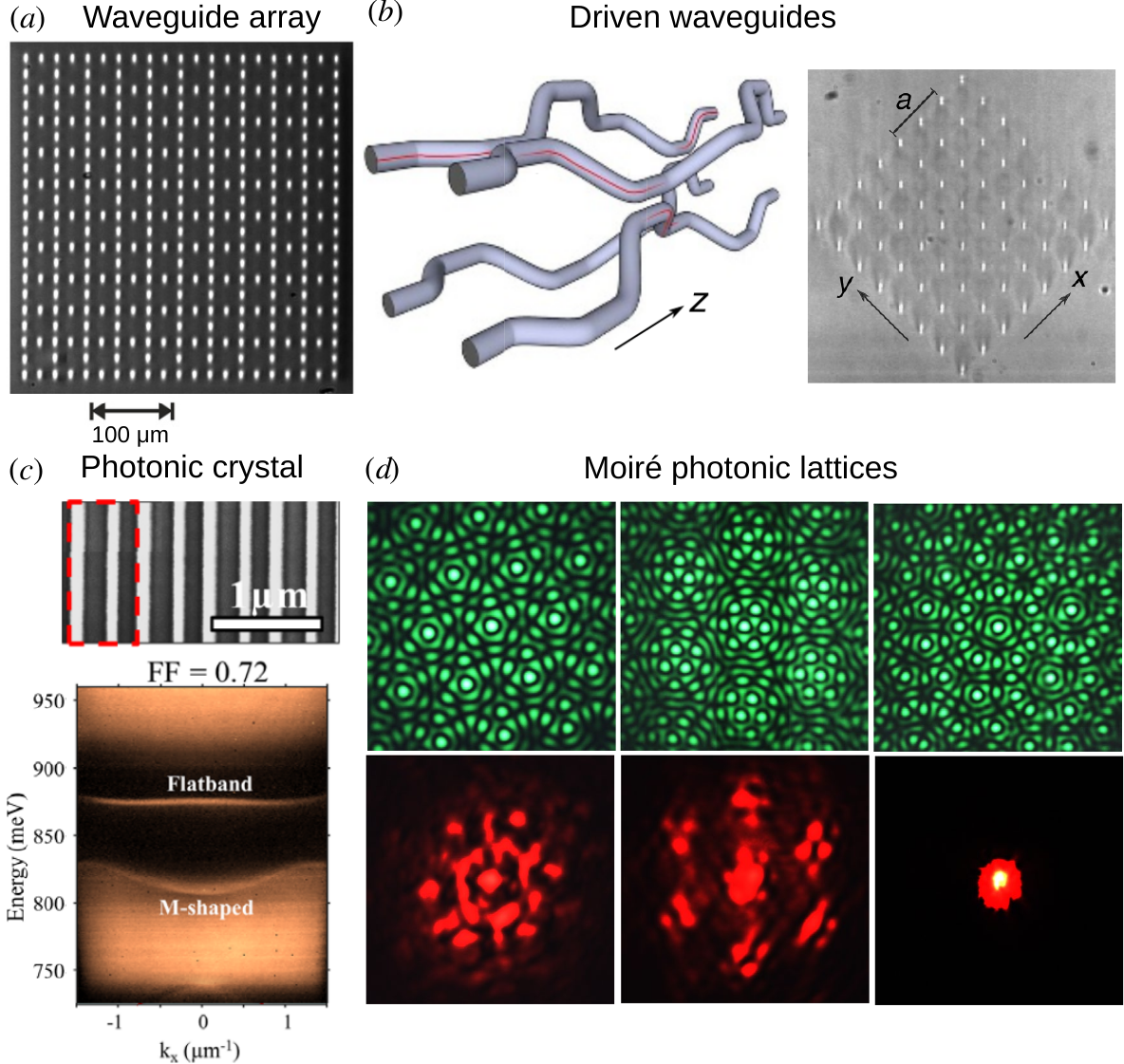}
    \caption{
        Examples of platforms for photonic flat bands.
        (a) Laser-written waveguide array implementing the 2D Lieb lattice, adapted from Ref.~\cite{guzman2014experimental}.
        (b) Longitudinally-modulated waveguide array implementing a Floquet flat band, adapted from Ref.~\cite{mukherjee2017experimental}.
        (c) 1D photonic crystal exhibiting a flat band at a fine-tuned value of its filling factor (FF) and its angle-resolved reflectivity spectrum, adapted from Ref.~\cite{nguyen2018symmetry}.
        (d) Optically-induced moir\'{e} photonic lattices (top row) revealing twist angle-controlled localization of a probe beam (bottom row), adapted from Ref.~\cite{wang2020localization}.
    }
    \label{fig:photonic_FBs}
\end{figure}

Optical waveguide arrays have formed a flexible platform for realizing many of the above fine-tuning effects. 
Waveguide arrays lend themselves naturally to implementation of flat bands thanks to the ability to precisely control the waveguide positions and thereby engineer the requisite fine-tuned coupling strengths and network geometries~\cite{li2008systematic}.
A variety of flat bands has been implemented in experiments using waveguide arrays including the sawtooth chain~\cite{weimann2016transport}, stub chain~\cite{rojas-rojas2017quantum,real2017flat}, diamond chain~\cite{mukherjee2015observation1,rojas-rojas2017quantum}, Sierpinsky gasket~\cite{travkin2017compact,hanafi2022localized}, Lieb lattice~\cite{guzman2014experimental,vicencio2015observation,mukherjee2015observation,yu2020isolated} (shown in Fig.~\ref{fig:photonic_FBs}(a)), as well as lattices hosting nontrivial loop boundary states caused by band touchings~\cite{tang2020photonic,gao2022demonstration}.

Introducing a transverse or longitudinal modulation to the waveguides can be used to implement various driving perturbation terms to flat band lattices. 
For example, the flat band in the diamond chain has been shown to be robust in the presence of external driving along the lattice axis~\cite{mukherjee2017observation}. 
Fine-tuning of periodic longitudinal driving in square lattices can create Floquet flat bands with non-trivial topology~\cite{maczewsky2017observation,mukherjee2017experimental,afzal2020realization}, shown in Fig.~\ref{fig:photonic_FBs}(b), 
and non-abelian Thouless pumping of spatially compact excitations~\cite{Sun2022nonabelian}. 
Transverse driving has been used to observe Bloch oscillations in photonic rhombic lattices~\cite{xia2020observation}.
%Inverse Anderson transition has been implemented in cold atoms for the diamond chain~\cite{li2022aharonov}.

%   Figure 11

Typically in nanophotonic systems fine-tuning is an essential ingredient for the realization of flat bands, owing to the inevitable presence of long-range couplings that cannot be precisely controlled. 
For example, fine-tuning of the radius of dielectric rods in 2D photonic crystals was used to obtain an accidental degeneracy similar to the flat band of the Lieb lattice, first at microwave frequencies~\cite{huang2011dirac} and subsequently in optical metamaterials~\cite{moitra2013realization}. 
The emergence of flat bands in fine-tuned photonic crystals can be understood semi-analytically using perturbation theory in Fourier space by taking into account scattering between different wavevectors induced by the periodic refractive index modulation combined with symmetry-breaking perturbations~\cite{nguyen2018symmetry,munley2023visible}.
An example of a fine-tuned photonic crystal flat band is shown in Fig.~\ref{fig:photonic_FBs}(c).
Modifications to this perturbation theory~\cite{lou2021theory,dong2021flat,tang2021modeling} have enabled the description of flat bands in photonic moir\'{e} superlattices~\cite{sunku2018photonic,wang2020localization,mao2021magic,Yi2022,nguyen2022magic,du2023moire}, shown in Fig.~\ref{fig:photonic_FBs}(d), which are difficult to study using direct numerical simulations due to their large unit cell size.

Another inevitable ingredient of nanophotonic systems is losses - either through absorption or radiation to the far field, making this an ideal playground for exploring the physics of non-Hermitian flat bands. 
For example, the physics of non-Hermitian \(\mathcal{PT}\)-symmetric flat bands was explored using waveguide arrays with losses introduced by periodically inscribing strong scattering centres~\cite{tobias2019experimental} (Fig.~\ref{fig:FB_detuning}(c)). 
In the case of passive photonic crystals, the light cone marks a transition between bound and lossy modes that must be taken into account when fine-tuning degeneracies in the band structure~\cite{letartre2022analytical}. 
Mie theory provides a powerful semi-analytical approach for taking into account the interplay between local couplings and far-field radiation when engineering strong degeneracies of nanophotonic systems~\cite{hoang2022high}.

A recent experimental study on the border line between condensed matter and photonics - a Bose-Einstein condensate of \(~^{87}Rb\) atoms loaded into an optical momentum lattice supported by various laser beams - tested the impact of disorder on flat band states~\cite{zeng2024transition,mao2024transition}. The work reports on the realization of a sawtooth chain (coined Tasaki lattice in Ref.~\cite{zeng2024transition,mao2024transition}) and the observation of the transition from flat band localization (i.e. compact localized states) to Anderson localization similar to theoretical predictions discussed above.  

The interested reader can find more details on photonic flat bands in the recent reviews~\cite{leykam2018perspective,vicencio2021photonic}.

\section{Interactions}
\label{sec:V}

Flat bands and compact localized eigenstates are linear (single particle) concepts.
However, in photonics material-mediated photon-photon interactions naturally occur for strong optical power.
For many photons in the classical regime mean-field nonlinear terms appear in the governing equations~\cite{kivshar2003optical}.
These nonlinear terms extend the dynamical regimes way beyond the Bloch band structure based one, allowing even for thermalization and chaos. 
Likewise in quantum interacting regimes, also accessible with photonic platforms~\cite{hartmann2016quantum,Noh2017quantum}, in general single particle CLS are replaced by extended many-body states. 
The impact that photonic interactions may have on flat bands including the unconventional phenomena they may unlock depend on the chosen lattice. 
From a fine-tuning standpoint, this quest for unusual phases can be rephrased as
(i) find sub-families of flat band networks for given interactions, or
(ii) tune special interactions to yield certain phenomena upon sub-families of flat band networks.

\subsection{Nonlinear regime}
\label{sec:V_1}

Destructive interference -- the underlying principle behind the existence of CLS -- is a concept which is not restricted only to the linear regime of a lattice. 
Hence, the fine-tuning approach for generating lattices supporting CLS and their flat bands can be extended to generating nonlinear lattices which support strictly compact photonic excitations. 
%In this case, nonlinearity adds additional knobs to the generator schemes. 
Such compact solutions can be considered as extensions of linear CLS into the nonlinear regime, and they are special cases of a broad class of spatially localized time-periodic solutions of nonlinear lattices called {\it discrete breathers}~\cite{ovchinikov1970localized,sievers1988intrinsic,mackay1994proof,flach1998discrete,campbell2004localizing,flach2008discrete}. 
These solutions -- also known as {\it intrinsic localized modes} -- have been studied in the realm of nonlinear optics~\cite{lederer2008discrete} and they have been experimentally observed e.g. with waveguide arrays~\cite{eisenberg1998discrete,fleischer2003observation} and optical fibers~\cite{solli2007optical}. 

%   Figure 12
\begin{figure}[h!]
    \centering
    \includegraphics[width=0.95\columnwidth]{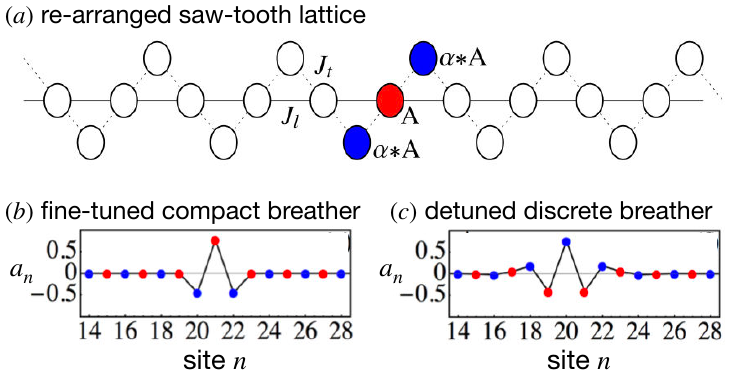}
    \caption{
        (a) Saw-tooth lattice with Kerr nonlinearity.
        The CLS is colored in blue and red, where \(\alpha = -J_l/J_t\). 
        (b) Spatial profile of a compact breather fine tuned at specific strengths of nonlinearity and hopping terms.  
        (c) Spatial profile of a non compact discrete breather at slightly detuned nonlinear and hopping strengths with respect to (b). 
        Adapted from Ref.~\cite{johansson2015compactification}. 
    }
    \label{fig:CB_tuning}
\end{figure} 

Strictly compact breathers harvesting on destructive interference have then been found in several flat band models in presence of Kerr nonlinearity in the past decade, from the Kagome to the diamond chain~\cite{vicencio2013discrete,johansson2015compactification,gligoric2016nonlinear,belicev2017localized,real2018controlled,danieli2018compact}. 
One of these sample cases is the nonlinear saw-tooth chain~\cite{johansson2015compactification} shown in Fig.~\ref{fig:CB_tuning}(a).
This unfolded version of Fig.~\ref{fig:fb-ex}(b) is an arrangement which facilitates the experimental realization by drastically reducing longer range hoppings~\cite{Weimann2016transport}.
The CLS profile is shown with blue and red dots.
The amplitudes in the blue sites are rescaled by a prefactor \(\alpha\) with respect to the amplitude in the red site, with \(\alpha\) depending on the hopping strengths of the lattice. 
In presence of a Kerr nonlinearity, the compact breather (Fig.~\ref{fig:CB_tuning}(b)) continues to exist at a specific fine-tuned condition between the nonlinear strength and the hopping strengths.
However, when this condition is slightly violated, the compact breather gains exponential tails and turns into a regular discrete breather (Fig.~\ref{fig:CB_tuning}(c)). 
In other words, compact breathers of the nonlinear saw-tooth lattice in Fig.~\ref{fig:CB_tuning}(a) are special fine-tuned cases of continuous families of exponentially localized discrete breathers. 

%   Figure 13
\begin{figure}[h!]
    \centering
    \includegraphics[width=0.95\columnwidth]{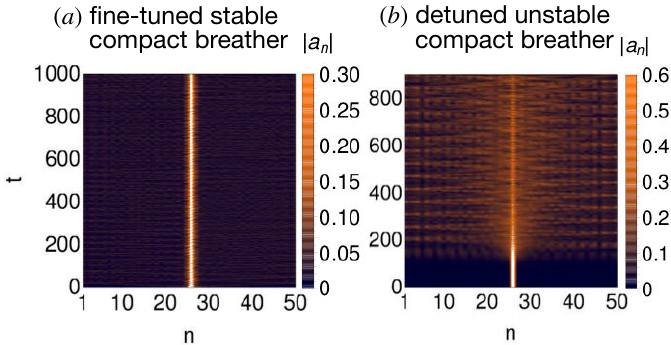}
    \caption{
        Stable (a) and unstable (b) propagation of an initially excited compact breather in a cross-stitch lattice with Kerr nonlinearity for two different nonlinear strengths. 
        Adapted from Ref.~\cite{danieli2018compact}.  
    }
    \label{fig:CB_tevo}
\end{figure}

The compact breather fine-tuning for the sawtooth lattice is due to the linear CLS of the model having different amplitudes at different sites.  
%is not the only possible scenario occurring in flat band networks with Kerr nonlinearity. 
At variance, the linear CLS of the cross-stitch lattice (Fig.~\ref{fig:fb-ex}(a)) can be continued as a compact breather to any nonlinear strength.
In this case, the Kerr term parametrically shifts the CLS frequency while keeping its spatial profile intact~\cite{danieli2018compact}. 
The reason is the homogeneity of the linear CLS.
Indeed, in the cross-stitch all non-zero CLS components have same amplitude up to phase factors, while this condition in the CLS of the saw-tooth is violated by the prefactor \(\alpha\). 
This fact has been formalized into a criterium stating that for Kerr nonlinearity any homogeneous CLS can be continued as compact breathers in the nonlinear regime, with the nonlinear strength parametrically shifting their frequency~\cite{danieli2018compact}.
On the other hand, non-homogeneous CLS can be continued as discrete breathers in the nonlinear regime of a flat band lattice, and compactness can be accidentally re-obtained at fine-tuning conditions involving the nonlinear strength and hopping parameters. 

Building upon these results and sample cases, the real space flat band generator~\cite{maimaiti2017compact,maimaiti2019universal,maimaiti2021flatband} discussed in Sec.~\ref{sec:IV_1} has been adapted to the nonlinear case and turned into systematic compact breathers generator~\cite{danieli2021compact}. Such schemes do not assume the homogeneity condition of compact breathers or even the presence of any flat band in the linear system, and can be extended to considerations of fine-tuned nonlocal nonlinear terms~\cite{danieli2021compact}. 
Together with their existence criteria, it has been found that linear instability of compact breathers arises from resonance mechanism between the breather itself and linear oscillations of the underlying lattice~\cite{johansson2015compactification,danieli2018compact}. 
In short, their frequency should be tuned away from the dispersive part of the spectrum of the linear flat band system.
This is visualized in Fig.~\ref{fig:CB_tevo}, where a slightly perturbed compact breather of the cross-stitch is tuned to two different frequencies via the nonlinear strength.
A stable case where resonances are avoided is shown in Fig.~\ref{fig:CB_tevo}(a), while an unstable one is shown in Fig.~\ref{fig:CB_tevo}(b) where the breather frequency is tuned to resonate with linear dispersive modes.  

The impact of compact breathers on the dynamics of nonlinear flat band networks is still a rather unexplored yet promising research area.
For instance, it has been shown that compact breathers can turn into tunable nonlinear Fano resonances~\cite{miroschnichenko2010fano} for the propagation of light (a mechanism also ignited by onsite impurities in flat bands)~\cite{ramachandran2018fano}. 
It is also possible to use nonlinear flat bands with non-Hermitian impurities to engineer coherent perfect absorbers (also known time-reversed laser)~\cite{danieli2020casting}.
Furthermore, nonlinear flat band lattices can also allow control over the mobility of discrete solitons, as it has been demonstrated in the nonlinear Kagome~\cite{vicencio2013discrete} and nonlinear Lieb lattice~\cite{real2018controlled}.  

All-band-flat networks demonstrate the highest sensitivity to nonlinearity. Since the linear part lacks any transport, nonlinear perturbations will in general restore transport. Surprisingly, in the nonlinear diamond chain with fine-tuned magnetic flux (Fig.~\ref{fig:fb_abf}), any spatially compact initial excitation remains therein confined along the propagation. In other words, Kerr nonlinearity does not destroy linear caging for that system~\cite{gligoric2019nonlinear,diliberto2019nonlinear}.
%However, it turned out that nonlinear caging requires fine-tuning. 
Indeed, generic nonlinear interactions including Kerr nonlinearity upon the manifold of all-bands-flat networks destroy caging and induce transport~\cite{danieli2021nonlinear}. 
Caging then is restored for fine-tuned families of ABF linear lattices.  
An example is shown in Fig.~\ref{fig:nonlinear_caging}(a,b) for two cases of the family of 1D two bands ABF networks with Kerr nonlinearity.
A fine-tuned lattice in Fig.~\ref{fig:nonlinear_caging}(a) (which correspond to the Creutz ladder~\cite{creutz1999end}) retains strict nonlinear caging as an initially compact excitation remains therein localized.
On the other hand, in a detuned ABF lattice in Fig.~\ref{fig:nonlinear_caging}(b) the initially compact excitation is propagating into the chain, confirming that caging is lost.
We also mention the impact of nonlinearity on momentum space linear Bloch modes which results in bifurcations and instabilities and the formation of localized states~\cite{Chang2021nonlinear}.

\subsection{Quantum regime -- a perspective}
\label{sec:V_2}

The quantized analog of the Kerr nonlinearity is the two-body interaction containing a product of two creation and two annihilation operators.
Its impact has been intensively studied in recent years in flat band networks from a mathematical and applied condensed matter perspective. 
The challenge of making photons interact results in interaction induced via {\it e.g.} doped media~\cite{hartmann2016quantum,Noh2017quantum} or through the optical nonlinear terms at the level of individual photons~\cite{Chang2014quantum}.
Most studies focused on finite-particle density regimes ({\it i.e.} where the ratio between the number of particle and the lattice size is constant). 
%Since photons are bosons, they can occupy the same single particle state. Occupation numbers of five or more photons per single particle state reduce most of the physics back to the nonlinear description. 
This has been a prolific research direction, as over the years phenomena of condensation and superconductivity have been discovered~\cite{zhao2012quantum,takayoshi2013phase,tsai2015interaction,metcalf2016matter,bercx2017spontaneous}, together with phenomena of absence of charge transport known as many-body flat band localization~\cite{danieli2020many,vakulchyk2021heat,kuno2020flat}, disorder-free localization~\cite{orito2021nonthermalized,danieli2022many}, magnon crystallization~\cite{Schulenburg2002macroscopic,Okuma2019series,Schnack2020magnon} and quantum many-body scars~\cite{hart2020compact,kuno2020flat_qs,tilleke2020nearest,danieli2021quantum,kuno2021multiple,kuno2020interaction}.
Some of these works harvested on the real space generator scheme for single particle lattices to highlight fine-tuned manifolds of interacting flat band networks exhibiting many-body flat band localization and quantum scars~\cite{danieli2021quantum,danieli2020many,danieli2022many}.

%   Figure 14
\begin{figure}[h!]
    \centering
    \includegraphics[width=\columnwidth]{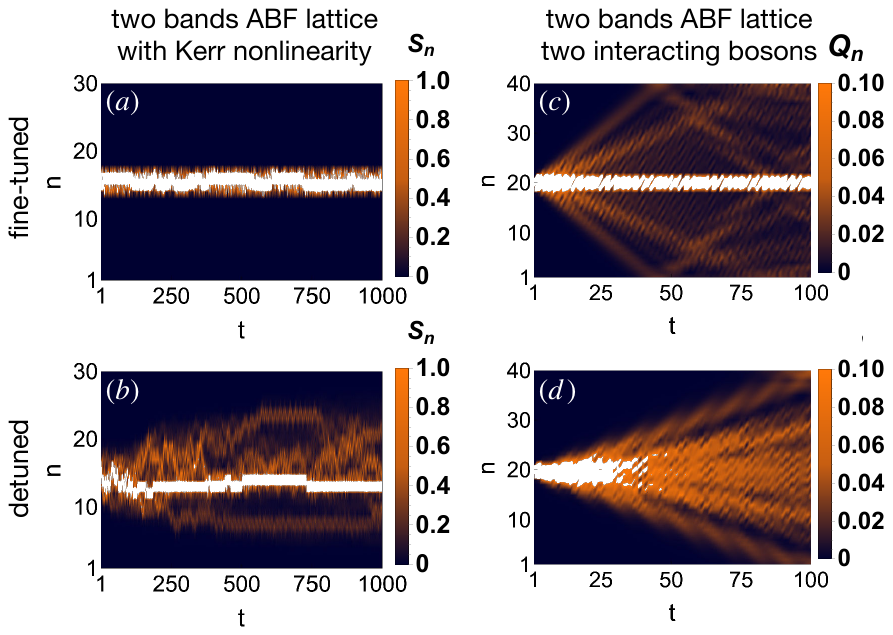}
    \caption{
        Interaction in \(\nu=2\) ABF lattices. 
        (a-b) Time evolution of nonlinear spatially compact excitations for a fine-tuned (a) and non fine-tuned case (b). 
        (c-d) Same as (a-b) for two interacting bosons. 
        Adapted from Ref.~\cite{danieli2021nonlinear,danieli2021quantum}
%        \textbf{Source: Fig.3 from PRB 104, 085131 (2021), and Figs.2 \& 3 from PRB 104, 085132 (2021)}
    }
    \label{fig:nonlinear_caging}
\end{figure}

Since photons are bosons, they can occupy the same single particle state.
As a rule of the thumb occupation numbers of five or more photons per single particle state reduce most of the physics back to the nonlinear description. 
Scaling down from the finite particle density regimes to fixed number of particles, higher dimensional linear photonic networks have been employed to study few quantum interacting particles in lattices.
For instance, the dynamics of two interacting bosons in one-dimensional disordered chains has been probed via two-dimensional photonic lattices, where the Hubbard interaction turns into an onsite diagonal term~\cite{Lee2014probing} -- an approach which has been tested experimentally~\cite{Xing2021anderson} and also employed for three interacting bosons~\cite{Yusipov2017quantum}. 
In the context of flat bands, this approach has been used to study the dynamics of two interacting bosons in 1D two band ABF networks with Bose-Hubbard interaction~\cite{danieli2021quantum}, shown in Fig.~\ref{fig:nonlinear_caging}(c,d). 
The propagation of two bosons initially located in the same unit-cell is obtained for the same two lattices used to visualize the nonlinear caging [fine-tuned network in Fig.~\ref{fig:nonlinear_caging}(a)] and its breaking [detuned network in Fig.~\ref{fig:nonlinear_caging}(b)]. 
In both lattices, a part of the probability distribution propagates ballistically, indicating the spreading of the particle pair along the networks. 
However, in the fine-tuned case Fig.~\ref{fig:nonlinear_caging}(c), a substantial fraction of the probability distribution remains frozen around the unit cell where both particles were initial located.
This frozen portion of the distribution is instead absent in the detuned case Fig.~\ref{fig:nonlinear_caging}(d).  
Such phenomenon is due to overlap between the initial condition and energy-renormalized two particles caged states, whose existence is ensured by the very same fine-tuning condition that guarantee nonlinear caging in the classical nonlinear regime~\cite{danieli2021quantum}. 
These two-particles compact states in ABF network can be found also for larger finite number of particles, and they generalize the single particle CLS~\cite{danieli2021quantum}.
Furthermore, such many-body CLS can be fine-tuned in networks which host a mix of flat and dispersive bands~\cite{hart2020compact,kuno2020flat_qs,tilleke2020nearest,kuno2021multiple,swaminathan2023signatures}. 
However, the challenge of turning them into ``localized quantum states of light'' in photonic lattices (as dubbed in Ref.~\cite{rojas-rojas2017quantum}) remains to this date an exciting research direction. 
%Likewise, this challenge may be extended toward photonic disorder-free localization of light, by devising ABF lattices with fine-tuned mediated density-density (?photon-photon?) interaction which completely vanish their transport~\cite{danieli2020many,vakulchyk2021heat,kuno2020flat}.
The utilization of quantum many-body physics of interacting photons~\cite{Chang2014quantum,hartmann2016quantum,Noh2017quantum,rojas-rojas2017quantum} for flat band networks is still an avenue to be explored.

%\begin{enumerate}

%\item  following the first evolution cycle in the late eighties~\cite{kohmoto1986electronic1,sutherland1986localization,lieb1989two,mielke1991ferromagnetism,tasaki1992ferromagnetism}, quantum interaction in flat band focused on finite-density many body regimes typical of condensed matter physics\\ mention:
%condensation \& superconductivity~\cite{zhao2012quantum,takayoshi2013phase,tsai2015interaction,metcalf2016matter,bercx2017spontaneous, ....}, quantum scars~\cite{hart2020compact,kuno2020flat_qs,danieli2021quantum,kuno2021multiple,kuno2020interaction}, many-body flatband localization~\cite{danieli2020many,vakulchyk2021heat,kuno2020flat} and ergodicity breaking~\cite{orito2021nonthermalized,danieli2022many} where fine-tuning schemes were applied~\cite{danieli2021quantum,danieli2020many,danieli2022many}

%\item   
%Suitable photonic platforms range from emulating few quantum interacting particles with higher dimensional linear photonic networks to true quantum features of photonic fields interacting with quantum regime solid state devices - superconducting devices, Rydberg systems, and others ...

%\end{enumerate}

\subsection{Applications in photonics}
\label{sec:V_3}

% \begin{enumerate}
%     \item 
%     compact localized lines (planes), loops, zig-zags in 2D (3D) nonlinear flatband lattices
%     \alexei{AA: duplicates the single particle applications}
%     \item
%     engineer experimentally feasible nonlocal interaction terms
%     \alexei{AA: Carlo did you add this?}
%     \item
%     topological states in interacting ABF lattices -- from 1D sample networks~\cite{zurita2020topology,lang2021non} to crafting higher dimensional lattices 
%     \alexei{AA: I do not yet see the relevance of Ref.~\cite{lang2021non}}
%     \item 
%     effective Bose-Hubbard models in photonics~\cite{hartmann2016quantum} and citations therein
%     \item
%     perspective for future experimental results in photonics lattices 
% \end{enumerate}  

%Areas to cover: (1) Exciton-polaritons, (2) Nanolasers, (3) Kerr nonlinearity in waveguide arrays and optically-induced moir\'{e} structures, (4) Quantum simulation.

Interactions in photonics usually appear in the form of mean-field nonlinearities, arising due to the Kerr effect or saturable gain.
Thus, most experiments studying the interplay between interactions and flat bands have focused on this limit, with the aim of exploring novel paradigms for the design and control of lasers and other nonlinear optical devices.

\begin{figure}[h!]
    \centering
    \includegraphics[width=\columnwidth]{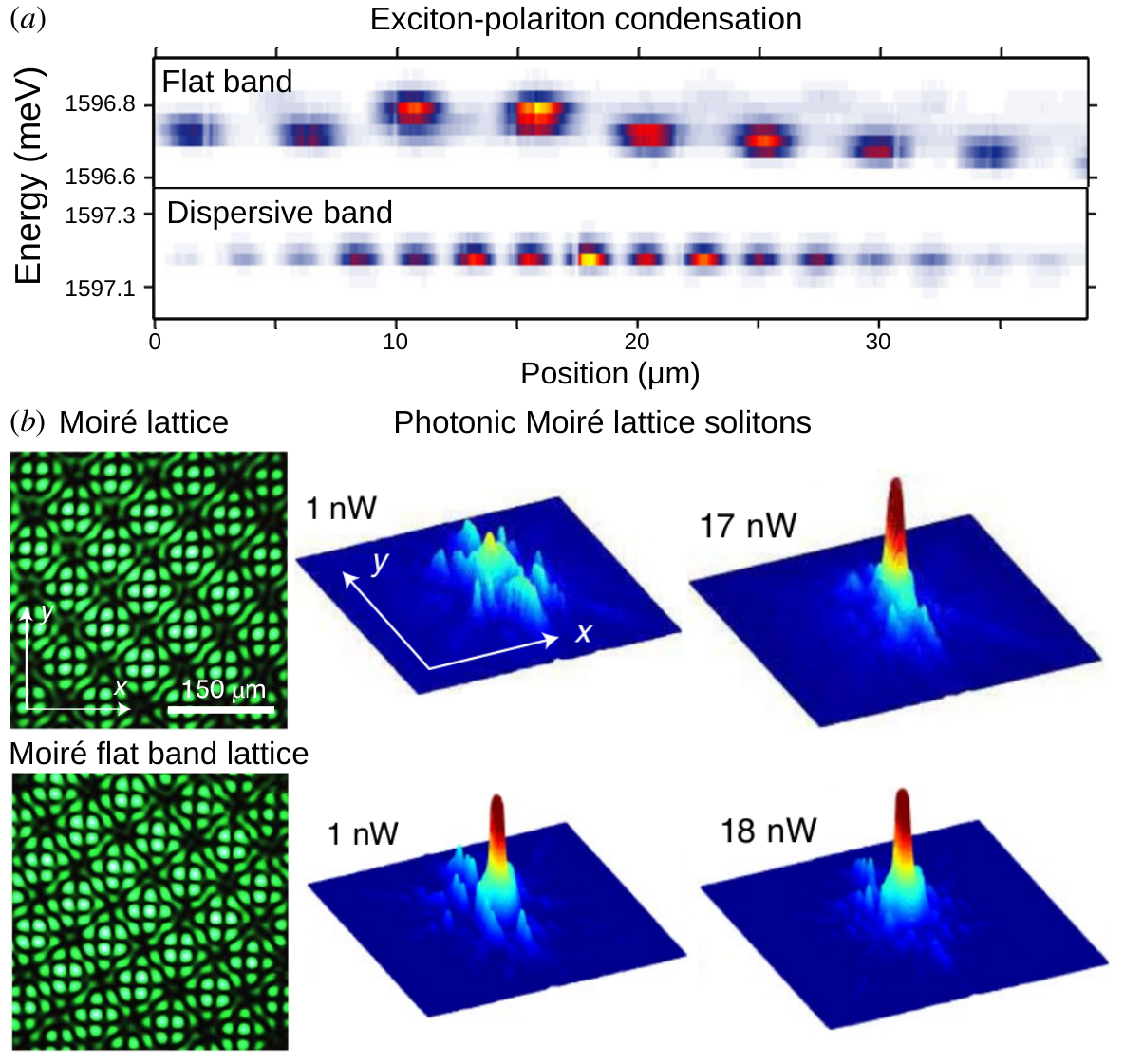}
    \caption{
        Photonics experiments combining flat bands with nonlinearity.
        (a) Spatially-resolved emission spectra from flat band and dispersive band exciton-polariton condensates, adapted from Ref.~\cite{baboux2016bosonic}.
        The former exhibits fragmentation into many incoherent condensates due to disorder-induced condensates, while the lattice exhibits coherence over many lattice sites.
        (b) Soliton formation in a optically-induced photonic moir\'{e} lattices, adapted from Ref.~\cite{Fu2020}.
        For a twist angle corresponding to a flat band, strong localization is observed for both low power (linear) and high power (nonlinear) probe beams. 
        Tuning away from this critical twist angle yields strong diffraction in the linear limit.
    }
    \label{fig:photonic_interactions}
\end{figure}

Exciton-polariton condensates in structured microcavities are a particularly valuable platform, given that they support strong repulsive Kerr nonlinearity (inherited from their excitonic component) and fine-tuned gain and loss landscapes can be created by adjusting the spatial profile of the external pump laser~\cite{AMO2016934}.
For example, Baboux et al~\cite{baboux2016bosonic} observed exciton-polariton condensation in a quasi-one-dimensional stub lattice of micropillars, finding that the flat band condensates become fragmented with a low coherence length due to the intrinsic fabrication disorder, preventing phase-locking between different compact localized states, as shown in Fig.~\ref{fig:photonic_interactions}(a).
These observations suggest a need to implement controlled perturbations (stronger than any underlying disorder) to observe flat band-derived condensates or lasing with non-trivial long-ranged correlations.
A more recent study using the exciton-polariton platform has illustrated how the driven-dissipative dynamics of media with saturable gain can be surprisingly counter-intuitive;
novel classes of localized modes can be created by engineering of the pump spatial profile and phase~\cite{jamadi2022reconfigurable} even when the underlying lattice does not host any flat bands.

Another experimental platform promising for the study of interactions in flat bands which has emerged in the last few years is nonlinear waveguide arrays. 
For example, even though the ambient Kerr nonlinearity of laser-written waveguide arrays is relatively weak, sufficiently high intensities leading to the formation of discrete solitons~\cite{Szameit:05} can be achieved using pulsed lasers.
Recent experiments have observed bulk and edge solitons in complex two-dimensional driven lattices~\cite{mukherjee2020observation,doi:10.1126/science.abd2033} that can be fine-tuned to obtain Floquet flat bands~\cite{maczewsky2017observation,mukherjee2017experimental}. 

%   Figure 15

One complication with studying interactions in the flat band limit using pulsed lasers is the finite bandwidth of the probe laser combined with the inevitable frequency dispersion of the coupling coefficients;
even if one has fine-tuned the lattice parameters to obtain a flat band at the central frequency of the probe, the spatial dispersion for other frequencies will be nonzero,
leading to diffraction and making it difficult to unambiguously observe nonlinearity-induced delocalization of the flat band state.
A potential solution to this problem is to use more sophisticated Floquet driving schemes to minimize the frequency dispersion of the coupling coefficients~\cite{Szameit2009}, in effect creating a flat band in the frequency domain~\cite{10.1063/5.0059525}. 

Another alternative is to use optically-induced waveguides photorefractive materials, which support both fine-tuning and the ability to observe nonlinear effects using a continuous wave probe laser~\cite{tang2020photonic},
which has enabled the recent observation of solitons in moir\'{e} superlattices~\cite{Fu2020}, shown in Fig.~\ref{fig:photonic_interactions}(b).

Turning to photonic crystals and metasurfaces, the integration of optical gain media into these structures is enabling the development of novel lasers based on concepts from topological photonics~\cite{Smirnova2020,Yang2022,Contractor2022,Ma2023,OtaTakataOzawaAmoJiaKanteNotomiArakawaIwamoto+2020+547+567}. 
While these studies largely focus on designs insensitive to perturbations, owing to the presence of difficult-to-control long-range couplings,
these structures can also support compact localized resonances and flat bands arising from the interplay between radiation losses and fine-tuning of the structural parameters~\cite{sgrignuoli2019compact,murpetit2020van,Mao2021,hoang2022high,hoang2023photonic}.
Beyond lasing, flat bands in photonic crystals show promise as a means of enhancing light-matter interactions~\cite{yang2023photonic}.

Photonic lattices can also be used to simulate different quantum systems, including interacting flat band models -- known as {\it photonic quantum simulators}~\cite{Aspuru-Guzik2012},
either using networks of optical cavities with ultrastrong nonlinearities~\cite{hartmann2016quantum} or by mapping the system of interest into a higher dimensional non-interacting system.
The latter has been proposed for the Creutz ladder with Hubbard-like interaction~\cite{zurita2020topology} and demonstrated experimentally for the one-dimensional Hubbard model~\cite{Olekhno2020}, paving the way to probe many interesting phenomena related to the interplay of topology and interactions. 
Another useful technique in this direction is the ability to emulate high-dimensional lattices with tailored long-range interactions using time modulation and the concept of synthetic dimensions~\cite{PhysRevLett.130.143801},
enabling the recent emulation of a sawtooth flat band lattice using two coupled ring resonators~\cite{10.1117/1.AP.4.3.036002}.

A recent theoretical study showed that flat band induced trapping of excitations in a nonlinear ABF diamond chain is destroyed on the quantum level of two bosons and quickly recovered when increasing the bosonic occupation numbers~\cite{kolovsky_2023}.
In parallel Martinez et al realized the setup using superconducting qubit arrays~\cite{martinez2023interaction}.
The experiments demonstrated that noninteacting photons are blocked due to the underlying ABF network, while the controlled adding of interaction lead to a transport of a pair of two interacting photons~\cite{martinez2023interaction}.

\section{Summary and outlook}	
\label{sec:conclusions}

% \begin{enumerate}
%     \item Construction and fine-tuning -- flatbands form continuous manifolds
%     \item Flatbands react differently to perturbations producing interesting effects
%     \item Further fine-tuning is possible: nonperturbative delocalization, many-bobdy scars and nonergodicity
% \end{enumerate}  

flat bands provide a fascinating testbed for novel and interesting phases of matter, that emerge in perturbed flat bands.
The challenge is that flat bands require symmetries or fine-tuning to engineer them that often are far from obvious.
One of the main results in the construction of flat bands is the not so obvious degree of flat band fine-tuning available.
Despite flat bands systems being fine-tuned or protected by symmetry and therefore a priori fragile to perturbations, they form continuous manifolds in the space of Hamiltonian parameters, and many translationally invariant perturbations preserve the flatness.
However, most physically relevant perturbations move the Hamiltonian away from that flat band manifold.
Furthermore, perturbation can break the flatness of a band or allow for further fine-tuning allowing to preserve the flatness, as happens with interactions where combining flat band lattices and interactions with the right degree of fine-tuning one can achieve nonergodic behaviour and suppress of some or all of the transport.

flat bands react differently to perturbations allowing to engineer many interesting phenomena, including nonperturbative transitions.
Orthogonal flat bands, e.g. with orthogonal CLS, are especially convenient for analysis.
The effect of weak perturbations can be understood efficiently by projecting the perturbation onto the flat band resulting in an {\it effective} Hamiltonian.
Orthogonality ensures that the projector onto the band is local, and therefore so is the effective projected model.
These effective projected models can sometimes be mapped onto the known models, or provide helpful insights into the properties of the perturbed flat bands.
Interactions further complicate the picture, however, further fine-tuning is possible which leads to emergent conservation laws that trap some particles, producing quantum caging, or suppress transport producing non-ergodic many-body systems.

Despite all the recent progress, we are confident that the best of flat band research still lies ahead of us, with numerous open questions to be addressed. 
%flat bands are known to exist in non-translationally invariant settings, like systems with d.c. fields, driven systems or quasicrystals.
%Understanding flatbands in these settings and their systematic generation is an interesting open problem.
%A related issue are the flatbands in hyperbolic geometry.

\section*{acknowledgement}
    CD and SF are grateful to the hospitality and support of the Visitors Program of the Max Planck Institute for the Physics of Complex Systems (Dresden, Germany) where parts of this work were completed.
%We are grateful to ...
%    Please insert acknowledgments of the assistance of colleagues or similar notes of appreciation here.
%\end{acknowledgement}

\section*{funding}
    CD acknowledges financial support from the project PNRR MUR CN\_0000013 HPC. 
    AA, SF acknowledge the financial support from the Institute for Basic Science (IBS) in the Republic of Korea through the project IBS-R024-D1.
    DL acknowledges support from the National Research Foundation, Singapore and A*STAR under its CQT Bridging Grant.
%    Please insert information concerning research grant support here (institution and grant number). Please provide for each funder the funder’s DOI according to https://doi.crossref.org/funderNames?mode=list.

%\printbibliography[env=bibnumeric]

\normalem
\bibliography{meta}

\end{document}